\documentclass[paper=a4, fontsize=11pt]{scrartcl}
\usepackage[T1]{fontenc}
\usepackage{fourier}
\usepackage[english]{babel}								 
\usepackage{amsmath,amsfonts,amsthm, amssymb} % Math packages
\usepackage{graphicx}	
\usepackage[title,titletoc,page]{appendix}        
\usepackage{color}
\usepackage{cite}
\usepackage{caption}
\usepackage{subcaption}
\usepackage[ruled,vlined]{algorithm2e}

\numberwithin{equation}{section}		% Equationnumbering: section.eq#
\numberwithin{figure}{section}			% Figurenumbering: section.fig#
\numberwithin{table}{section}				% Tablenumbering: section.tab#

\newtheorem{defi}{Definition}[section]
\newtheorem{lem}{Lemma}[section]
\newtheorem{thm}{Theorem}[section]
\newtheorem{cor}[lem]{Corollary}
\newtheorem{remark}{Remark}[section]

\newtheorem{prop}{Proposition}[section]

\title{A Theory of Computational Resolution Limit for Line Spectral Estimation}
\author{
	Ping Liu\thanks{\footnotesize Department of Mathematics, 
		HKUST,  Clear Water Bay, Kowloon, Hong Kong (pliuah@connect.ust.hk).}
	\; and Hai Zhang\thanks{\footnotesize 
		Department of Mathematics, 
		HKUST,  Clear Water Bay, Kowloon, Hong Kong (haizhang@ust.hk). Hai Zhang was supported by Hong Kong RGC grant GRF 16306318 and GRF 16305419.}}

\begin{document}
	
	\maketitle

	\begin{center}
		\textbf{Abstract}
	\end{center}
	Line spectral estimation is a classical signal processing problem that aims to estimate the line spectra from their signal which is contaminated by deterministic or random noise. Despite a large body of research on this subject, the theoretical understanding of this problem is still elusive. In this paper, we introduce and quantitatively characterize the two resolution limits for the line spectral estimation problem under deterministic noise: one is the minimum separation distance between the line spectra that is required for exact detection of their number, and the other is the minimum separation distance between the line spectra that is required for a stable recovery of their supports. The quantitative results imply a phase transition phenomenon in each of the two recovery problems, and also the subtle difference between the two. We further propose a sweeping singular-value-thresholding algorithm for the number detection problem and conduct numerical experiments. The numerical results confirm the phase transition phenomenon in the number detection problem. 
	
	\medskip
	
	\textbf{Keywords}: Line spectral estimation, resolution limit, phase transition.  
	
	\section{Introduction}\label{introduction}
	%\subsection{}
	This paper is concerned with recovering the number and supports of a collection of line spectra from their contaminated signal, which is usually termed the line spectral estimation (LSE) problem. It is at the core of diverse research fields such as wireless communication, radar, sonar, seismology and astronomy, and has received significant attention over the years. While the LSE problem was usually cast as a statistical parameter estimation problem with random noise in the measurement, we are interested in the case of deterministic noise. To be more specific, we consider the following mathematical model. Let $\mu=\sum_{j=1}^{n}a_{j}\delta_{y_j}$  be a discrete measure, 
where $y_j \in \mathbb R,j=1,\cdots,n$, represent the supports of the line spectra and $a_j\in \mathbb C, j=1,\cdots,n$ their amplitudes. We assume that $|y_j|\leq d$ for all $j$.  We denote 
	\[
	m_{\min}=\min_{j=1,\cdots,n}|a_j|,
	\quad 	d_{\min}=\min_{p\neq j}| y_p-y_j|.
	\] 
We sample the Fourier transform of $\mu$ at $M$ equispaced points:
	\begin{equation}\label{equ:modelsetting1}
	\mathbf Y(\omega_q) = \mathcal F \mu (\omega_q) + \mathbf W(\omega_q)= \sum_{j=1}^{n}a_j e^{i y_j \omega_q} + \mathbf W(\omega_q), \quad 1\leq q \leq M, 
	\end{equation}
where $\omega_1=-\Omega, \omega_2=-\Omega+h,\cdots, \omega_{M}=\Omega$. Here $h=\frac{2\Omega}{M-1}$
is the sampling spacing and $\mathbf W(\omega_q)$'s are the noise.  We assume that $|\mathbf W(\omega_q)|< \sigma$ with $\sigma$ being the noise level.
Throughout, we assume that $M>2n$ and that $h\leq \frac{\pi}{d}$.  The latter assumption excludes the non-uniqueness of the line spectra due to shifts by multiples of $\frac{2\pi}{h}$. 
Denote
	\[ 
	\mathbf Y =(\mathbf Y(\omega_1), \cdots, \mathbf Y(\omega_M))^T, \quad  [\mu] =(\mathcal F \mu(\omega_1),\cdots,\mathcal F \mu(\omega_M))^T\quad  \text{and} \quad \mathbf W =(\mathbf W(\omega_1),\cdots,\mathbf W(\omega_M))^T. 
	\]
	Then the noisy measurement can be written in the following form 
	\[\mathbf Y=[\mu]+\mathbf W.\]
    The LSE problem we are interested in is to recover the discrete measure $\mu$ from the above noisy measurement $\mathbf Y$.  We note that the LSE problem is closely related to the deconvolution problem in imaging where the measurement is the convolution of point sources and a band-limited point spread function $f$.  
%    \mu*f(x)=\sum_{j=1}^{n}a_j f(t-y_j)$. 
More precisely, in the presence of additive noise $\epsilon(t)$, the measurement is 
   	\begin{equation} \label{eq-decon}
   	y(t)=\mu*f(t)+\epsilon(t)=\sum_{j=1}^{n}a_j f(t-y_j)+\epsilon(t).
   \end{equation}
   	By taking the Fourier transform on both sides, we obtain 
   	\begin{equation} \label{eq-deconv}
   	\mathcal{F} y(\omega)= \mathcal{F}f(\omega)\cdot \mathcal F \mu(\omega)+ \mathcal{F} \epsilon(\omega)= \mathcal{F}f(\omega)(\sum_{j=1}^n a_j e^{i y_j\omega})+ \mathcal{F} \epsilon(\omega),
   	\end{equation}
 which is reduced to the LSE problem. 
%Assuming $0<c_1\leq \mathcal{F}f(x)\leq c_2$, according to the above formulation, it is straightforward to derive similar upper and lower bounds for the computational resolution limits of both number detection and support recovery in the deconvolution problem. 

	\subsection{Literature review}
It is well-known, since Rayleigh's work \cite{rayleigh1879xxxi}, that two sources (or line spectra as are called in this paper) can be resolved if they are separated more than the Rayleigh limit (or Rayleigh length in some literature) $\frac{\pi}{\Omega}$. Although it is an empirical limit, the Rayleigh limit plays an important role in many source or line spectra resolving algorithms. For example, it is proved that TV minimization can exactly resolve off-the-grid sources from their noiseless low-frequency measurement if they are separated more than several Rayleigh limits \cite{candes2014towards}. See also \cite{candes2013super, fernandez2016super, fernandez2013support} for the related researches in the resolving ability of TV minimization. Other sparsity promoting optimization based algorithms such as the BLASSO \cite{azais2015spike, duval2015exact, poon2019} and the atomic norm minimization \cite{tang2013compressed, tang2014near,chi2020harnessing} can also provably recover the off-the-grid sources under a minimum separation of several Rayleigh limits or certain non-degeneracy condition. This minimum separation requirement is necessary for general source recovery \cite{duval2015exact, tang2015resolution, li2018approximate, poon2019}, but can be relaxed for positive sources \cite{denoyelle2017support, morgenshtern2016super, morgenshtern2020super}.
	% and the performances are nearly optimal.
	
As the source separation distance decreases and falls below the Rayleigh limit, it becomes increasingly difficult to resolve them from the noisy measurement. In such sub-Rayleigh regime, a variety of parametric methods, including Prony's method \cite{Prony-1795}, MUSIC \cite{schmidt1986multiple, stoica1989music}, ESPRIT \cite{roy1989esprit} and Matrix Pencil method \cite{hua1990matrix, hua1991svd} are shown to have favourable performance. In general, parametric methods require a priori
information of the model order (or the number of line spectra) and their 
performances depend on it sensitively \cite{hansen2018superfast}. 
We note that ESPRIT and MUSIC algorithms are analyzed recently in \cite{li2019super, li2018stable}. We also refer to \cite{batenkov2019super} for the numerical performance of the Matrix Pencil method.
%Except the Prony's method, these algebraic methods usually estimate the line spectra based on an eigenvalue estimation of the data matrix. Also, they all need an a-prior estimation of the model order (the number of line spectra). 

%In the statistical setting, the model order is usually determined by generic information theoretic criteria (e.g. AIC \cite{akaike1998information, akaike1974new, wax1985detection}, BIC/MDL \cite{schwarz1978estimating, rissanen1978modeling, wax1989detection}) or methods based on eigenvalues of the estimated signal covariance matrix (e.g. \cite{lawley1956tests, chen1991detection, he2010detecting, han2013improved}). While the performance of these algebraic methods depends sensitively on the a-priori estimate of the model order \cite{hansen2018superfast}, their advantage is also evident from their appealing performance in the super-resolution region. 

Despite much progress in the development of algorithms, the theoretical understanding of the resolution limit is still elusive. A particular puzzle is the gap between the physical (classical) resolution limit and the limit from a data processing point of view. Precisely, the empirical Rayleigh limit is based on presumed resolving capabilities of the human visual system and is not useful for data elaborately processed, see for instance \cite{papoulis1979improvement, den1997resolution}. In the last century, this resolution limit puzzle had drawn much attention 
and was investigated extensively from the perspective of statistical estimation and hypothesis testing, see for instance \cite{helstrom1964detection, helstrom1969detection, lucy1992statistical, lucy1992resolution, den1996model}. Most of the studies focus on the two-point resolution limit which is defined to be the minimum detectable distance between two point sources at a given signal-to-noise ratio (SNR). 
%
%In \cite{helstrom1969detection}, Helstrom formulated the
%%the field received by aperture from radiation of incoherent sources as circular gaussian processes and formulated the 
%detection of two identical sources as a hypothesis test problem, and studied the probability of detecting the correct source number. He expressed it by the decision level, the duration of the observation, the separation of sources, the band-width in the measurement and the energy-to-noise ratio. Moreover, using the Cram\'er-Rao inequality, he derived in \cite{helstrom1969detection, helstrom1970resolvability} the lower bounds on the mean-square error of unbiased estimators for the source positions, the distance between the sources, and the radiance values. In \cite{lucy1992statistical, lucy1992resolution}, employing an approximate statistical theory, Lucy analyzed the required number of detected photons (similar to the notion of SNR) for resolving by certain strategies a pair of identical point sources. The achievable resolution by a iterative deconvolution technique for image restoration was also investigated numerically, under a specified criterion that two sources are deemed resolved. In \cite{den1996model, den2002model}, using a model fitting theory, two-point resolution of imaging system was studied that the probability of resolution was computed based on the structural change of the stationary points of the likelihood function. 
Especially, in \cite{shahram2004imaging, shahram2004statistical, shahram2005resolvability}, by unifying and generalizing much of the literature on the topic which spanned the course of roughly four decades, the authors derived explicit formula for the minimum SNR that is required to discriminate two point sources separated by a distance smaller than the Rayleigh limit.  
%studied the achievable resolution of incoherent imaging in 1 and 2 dimensional cases. More specifically, they cast the detection of two sources as a hypothesis test and estimated the required SNR of exact detection for many cases, e.g., the case of symmetrically located sources with equal and known intensities, the case of asymmetrically located sources with unknown intensities and so on. 
It is shown that the required SNR is inversely proportional to a certain power of the source separation distance and the power is different from case to case. 

On the other hand, 
Donoho first addressed the resolution limit from the optimal recovery point of view \cite{donoho1992superresolution}. He considered measures supported on the lattice $\{k\Delta\}_{k=-\infty}^{+\infty}$ and regularized by the so-called ``Rayleigh index''. He showed that the minimax error for the amplitude recovery with noise level $\sigma$ scales like $SRF^\alpha \sigma$, where $SRF=\frac{1}{\Omega\Delta}$ is the super-resolution factor, and $\alpha$ is a parameter depends on the Rayleigh index. This result highlights the importance of sparsity and SNR in the ill-posedness of this inverse problem. Further discussed in \cite{demanet2015recoverability}, the authors considered the case of $n$-sparse signals supported on a grid and showed that the scaling of the noise level for the minimax error should be $SRF^{2n-1}$. See also similar results for the multi-clumps case in \cite{li2018stable, batenkov2020conditioning}. 
However, these works mostly deal with the grid setting and do not address the recovery of source supports. 

In \cite{Moitra:2015:SEF:2746539.2746561}, under the off-the-grid setting, using novel extremal functions, the author established a sharp phase transition for the amplitude and support recovery in the relation between cutoff frequency ($\Omega$) and separation distance ($\Delta$). 
Recently in \cite{akinshin2015accuracy} the authors derived the required SNR for a stable recovery of supports in the LSE problem. They further derived sharp minimax errors for the support and the amplitude recovery in \cite{batenkov2019super}. More precisely, they showed that for $\sigma \lessapprox (SRF)^{-2p+1}$, where $p$ is the number of nodes (or the line spectra as are called in this paper) which form a cluster of certain type, 
the minimax error rate for reconstruction of the clustered nodes is of the order $(SRF)^{2p-2} \frac{\sigma}{\Omega} $, while for recovering the corresponding amplitudes the rate is of the order $(SRF)^{2p-1}\sigma$.  Moreover, the corresponding minimax rates for the recovery of the non-clustered nodes and amplitudes are $\frac{\sigma}{\Omega}$ and $\sigma$ respectively.  In an earlier work \cite{liu2019computational} by the authors of the paper, the computational resolution limit was proposed for the deconvolution problem (\ref{eq-decon}). By working directly with the measurement in (\ref{eq-decon}) other than the Fourier data in (\ref{eq-deconv}), and employing a multipole expansion method, they showed that the resolution limit for number detection is bounded above by $\frac{C(d)}{\Omega}\left(\frac{\sigma}{\sigma_{\min}(s)m_{\min}}\right)^{\frac{1}{2n-2}}$, where $C(d)$ is a constant depending on the interval with size $d$ where all the $n$ sources are located and $\sigma_{\min}(s)$ is a positive number characterizing the correlation of the
multipoles used in the reconstruction. Similar result on the resolution limit for the support recovery was obtained as well. 
This paper deals with the LSE problem. It can be viewed as an investigation for the deconvolution problem with Fourier measurement. The results obtained herein improves the bounds in  \cite{liu2019computational} by getting rid of the factor $\sigma_{\min}(s)$. 

%They showed that 

%when $\sigma \gtrapprox (\Omega d_{\min})^{2n-1}$, the minimax error of the reconstruction for the closely spaced sources and for some off-cluster nodes scales as $SRF^{2n-2}\frac{\sigma}{\Omega}$ and $\frac{\sigma}{\Omega}$ respectively. 

%the authors in resolved the puzzle for the amplitude and support recovery in the off-the-grid setting. It is shown in \cite{akinshin2015accuracy} that the minimum separation distance required for a stable support recovery is of the order $O(\frac{1}{\Omega}\cdot (\frac{1}{SNR})^{\frac{1}{2n-1}})$. When the sources are separated beyond this distance, \cite{batenkov2019super} demonstrated that the minimax error of support recovery scales as $SRF^{2n-2}\frac{\sigma}{\Omega}$.

%
% using techniques of "Prony mapping" and "quantitative inverse function theorem", the authors in  resolved the puzzle for the amplitude and support recovery in the off-the-grid setting. It is shown in \cite{akinshin2015accuracy} that the minimum separation distance required for stable support recovery is of the order $O(\frac{1}{\Omega}\cdot (\frac{1}{SNR})^{\frac{1}{2n-1}})$. When the sources are separated beyond this distance, \cite{batenkov2019super} demonstrated that the minimax error of support recovery scales as $SRF^{2n-2}\frac{\sigma}{\Omega}$.   

\subsection{Main contribution}
In this paper, we investigate the LSE problem for a cluster of closely spaced line spectra in the off-the-grid setting with deterministic noise. 
The main contribution is a quantitative characterization of the resolution limit to the spectral number detection problem.  Accurate detection of the spectral number (or the model order in some literature) is an important step in the LSE problem and many parametric estimation methods require the spectral number as a priori information. But there is few theoretical result which addresses the issue when this number is greater than two. The results we derived in the paper seem to be the first in this direction to our knowledge. To resolve this issue, we introduce the computational resolution limit $\mathcal{D}_{num}$ for the detection of $n$ spectra (see Definition \ref{computresolutionlimit}) where $n$ can be an arbitrary integer greater than or equal to two, and derive the following sharp bounds:  
\begin{equation}\label{equ:numbcrlbounds}
	\frac{0.81e^{-\frac{3}{2}}}{\Omega}\Big(\frac{\sigma}{m_{\min}}\Big)^{\frac{1}{2n-2}}<\mathcal{D}_{num}\leq  \frac{4.4\pi e}{\Omega }\Big(\frac{\sigma}{m_{\min}}\Big)^{\frac{1}{2n-2}},
	\end{equation}
where $\frac{\sigma}{m_{\min}}$ is viewed as the inverse of the SNR. 
It follows that with deterministic noise, exact detection of the spectral number is possible when the minimum separation distance of line spectra $d_{\min}$ is greater than $\frac{4.4\pi e}{\Omega }\Big(\frac{\sigma}{m_{\min}}\Big)^{\frac{1}{2n-2}}$,  and impossible  without additional a priori information when $d_{\min}$ is less than $\frac{0.81e^{-\frac{3}{2}}}{\Omega}\Big(\frac{\sigma}{m_{\min}}\Big)^{\frac{1}{2n-2}}$. The quantitative characterization of the resolution limit $\mathcal{D}_{num}$ implies a phase transition phenomenon in the number detection problem. We further propose a sweeping singular-value-thresholding algorithm for the number detection and conduct numerical experiments which confirm the prediction (see Section 5). 
The main technique used to derive the bounds for the resolution limit is the approximation theory in Vandermonde space (see Section \ref{section:approxinvande}). The approximation theory was first introduced in the authors' paper \cite{liu2019computational} for  real Vandermonde vectors. We generalize the theory to the case of complex vectors in this paper.

Following the same line of argument for the number detection problem, we also consider the support recovery problem in LSE. We introduced 
the computational resolution limit $\mathcal D_{supp}$ for the support recovery (see Definition \ref{computresolutionlimit2}) and derive the following bounds:
	\begin{equation}\label{equ:suppcrlbounds}
	\frac{0.49e^{-\frac{3}{2}}}{\Omega} \Big(\frac{\sigma}{m_{\min}}\Big)^{\frac{1}{2n-1}}<\mathcal{D}_{supp}\leq \frac{5.88\pi e}{\Omega }\Big(\frac{\sigma}{m_{\min}}\Big)^{\frac{1}{2n-1}}.
	\end{equation}
As a consequence, the resolution limit $\mathcal D_{supp}$ is of the order $O(\Big(\frac{\sigma}{m_{\min}}\Big)^{\frac{1}{2n-1}})$. We also show that 
when the minimum separation distance exceeds the upper bound of $\mathcal D_{supp}$, the deviation of recovered supports to the ground truth scales as $SRF^{2n-2}\frac{\sigma}{\Omega m_{\min}}$. Such results were also reported in the closely related work \cite{batenkov2019super} under a more general setting where some of the line spectra (or nodes as is called therein) form a cluster while the rests are well separated. Their results are based on the analysis of "Prony mapping" and the "quantitative inverse function theorem", which is different from ours.

	\subsection{Organization of the paper}
	The paper is organized in the following way. Section 2 presents the main results to the LSE problem. Section 3 introduces the main technique that is used to prove the main results. The readers may skip this section in the first reading. Section 4 proves all the main results of Section 2. In Section 5, a sweeping singular-value-thresholding algorithm for the number detection is proposed and numerical experiments are conducted. Section 6 provides a conclusion. Finally, the appendix proves some inequalities that are used in the paper.

	\section{Main results}\label{section:Mainresults}
	We present our main results on the resolution limit for the LSE problem in this section. All the results shall be proved in Section \ref{section:mainresultsproof}.
	We consider the case when the line spectra $y_j, j=1,\cdots,n$ are tightly spaced and form a cluster. To be more specific, we define the interval 
$$
I(n,\Omega):=\Big[-\frac{(n-1)\pi}{2\Omega},  \frac{(n-1)\pi}{2\Omega}\Big],
$$ and assume that $y_j\in I(n,\Omega), 1\leq j\leq n$. Recall that the Raleigh limit is $\frac{\pi}{\Omega}$.
	For a discrete measure $\hat \mu=\sum_{j=1}^{k} \hat a_j \delta_{\hat y_j}$, we can only determine if it is a solution to the LSE problem by comparing the data $[\hat \mu]$ it generated with the measurement $\mathbf Y$. In this principle, we introduce the following concept of $\sigma$-admissible measure (see also the error set in \cite{batenkov2019super}). 
	
	\begin{defi}{\label{determinecriterion1}}
		Given measurement $\mathbf Y$, we say that $\hat \mu=\sum_{j=1}^{k} \hat a_j \delta_{\hat y_j}$ is a $\sigma$-admissible discrete measure of $\mathbf Y$ if
		\[||[\hat \mu]-\mathbf Y||_{\infty}< \sigma.\]
	\end{defi}
	
	The set of $\sigma$-admissible measures of $\mathbf Y$ characterizes all possible solutions to the LSE problem with the given measurement $\mathbf Y$. A good reconstruction algorithm should give a $\sigma$-admissible measure. If there exists one $\sigma$-admissible measure with less than $n$ supports, then one may detect less than $n$ spectra and miss the exact one if there is no additional a priori information. 
On the other hand, if all $\sigma$-admissible measures have at least $n$ supports, then one can determine the number $n$ correctly if one restricts to the sparsest admissible measures. 
This leads to the following definition of resolution limit to the number detection problem in LSE.
	\begin{defi}\label{computresolutionlimit}
		For measurement $\mathbf Y$ generated by $n$ line spectra $\sum_{j=1}^{n}a_{j}\delta_{y_j}$, the computational resolution limit to the number detection problem is defined as the smallest nonnegative number $\mathcal D_{num}$ such that if 
		\[
		\min_{p\neq j} |y_j-y_p| \geq \mathcal D_{num}
		\]
		then there does not exist any $\sigma$-admissible measure  with less than $n$ supports for $\mathbf Y$.
	\end{defi}
	
	The above resolution limit is termed ``computational resolution limit'' to be distinct from the classic Rayleigh limit. It depends crucially on the SNR and the sparsity of line spectra, in contrast to the latter which depends only on the available frequency band-width in the measurement. We now present sharp bounds for this computational resolution limit.
	
	\begin{thm}\label{upperboundnumberlimithm0}
		Let $\mathbf Y$ be a measurement generated by  $\mu =\sum_{j=1}^{n}a_j\delta_{y_j}$ which is supported on $I(n,\Omega)$. Let $n\geq 2$ and assume that  the following separation condition is satisfied 
		\begin{equation}\label{upperboundnumberlimithm0equ0}
		\min_{p\neq j}\Big|y_p-y_j\Big|\geq \frac{4.4\pi e}{\Omega }\Big(\frac{\sigma}{m_{\min}}\Big)^{\frac{1}{2n-2}}.
		\end{equation}
		Then there do not exist any $\sigma$-admissible measures of \,$\mathbf Y$ with less than $n$ supports.
	\end{thm}
	Theorem \ref{upperboundnumberlimithm0} gives an upper bound for the computational resolution limit $\mathcal D_{num}$. 
%With minimum separation distance greater than the upper bound in (\ref{upperboundnumberlimithm0equ0}), the number detection problem is regularized, and any algorithm targeting at the sparsest admissible measures can recover the correct number. 
Compared with Rayleigh limit $\frac{\pi}{\Omega}$, the upper bound indicates that resolving the number of the line spectra in the sub-Rayleigh regime is theoretically possible if the SNR is sufficiently large. We next show that the above upper bound is optimal.
	% by the following proposition that  $\mathcal D_{num} \geq \frac{0.7(\frac{1}{e})^{\frac{3}{2}}}{\Omega}\big(\frac{\sigma}{m_{\min}}\big)^{\frac{1}{2n-2}}$. 
	\begin{prop}\label{numberlowerboundthm0}
For given $0<\sigma<m_{\min}$ and integer $n\geq 2$, there exist  $\mu=\sum_{j=1}^{n}a_j\delta_{y_j}$ with $n$ supports, and $\hat \mu=\sum_{j=1}^{n-1}\hat a_j \delta_{\hat y_j}$ with $n-1$ supports such that 
$||[\hat \mu]-[\mu]||_{\infty}< \sigma$. Moreover
\[
 \min_{1\leq j\leq n}|a_j|= m_{\min}, \quad \min_{p\neq j}|y_p-y_j|= \frac{0.81e^{-\frac{3}{2}}}{\Omega}\Big(\frac{\sigma}{m_{\min}}\Big)^{\frac{1}{2n-2}}.
  \]
		
\end{prop}
	The above result gives a lower bound for the computational resolution limit $\mathcal D_{num}$ to the number detection problem. Combined with Theorem \ref{upperboundnumberlimithm0}, it reveals that the computational resolution limit $\mathcal D_{num}$ is of the order $O(\frac{(1/SNR)^{\frac{1}{2n-2}}}{\Omega})$.
We emphasize that similar to parallel results in \cite{akinshin2015accuracy, batenkov2019super,liu2019computational}, our bounds are the worst-case bounds, and one may achieve better bounds with high probability for the case of random noise.  

\medskip	
	We now consider the support recovery problem in the LSE problem. We first introduce the following concept of $\delta$-neighborhood of a discrete measure. 
	\begin{defi}\label{deltaneighborhood}
		Let  $\mu=\sum_{j=1}^{n}a_j \delta_{y_j}$ be a discrete measure and let $\delta>0$ be such that the $n$ intervals $(y_k- \delta, y_k + \delta), 1\leq k \leq n$ are pairwise disjoint. We say that 
		$\hat \mu=\sum_{j=1}^{n}\hat a_{j}\delta_{\hat y_j}$ is within $\delta$-neighborhood of $\mu$ if each $\hat y_j$ is contained in one and only one of the n intervals $(y_k- \delta, y_k + \delta), 1\leq k \leq n$. 
	\end{defi}
	
	According to the above definition, a measure in a $\delta$-neighbourhood preserves the inner structure of the real line spectra. For any stable support recovery algorithm, the output should be a measure in some $\delta$-neighborhood. Moreover, $\delta$ should tend to zero as the noise level $\sigma$ tends to zero.  We now introduce the computational resolution limit for stable support recovery. For ease of exposition, we only consider measures supported in $I(n, \Omega)$ where $n$ is the number of supports. 
	
	\begin{defi}\label{computresolutionlimit2}
		For measurement $\mathbf Y$ generated by $\mu=\sum_{j=1}^{n}a_j \delta_{y_j}$ which is supported in $I(n, \Omega)$, the computational resolution limit to the stable support recovery problem is defined as the smallest nonnegative number $\mathcal D_{supp}$ so that if
\[
		\min_{p\neq j, 1\leq p,j \leq n} |y_p-y_j|\geq \mathcal{D}_{supp},
\]  
then there exists $\delta>0$ such that any $\sigma$-admissible measure for $\mathbf Y$ with $n$ supports in $I(n, \Omega)$ is within $\delta$-neighbourhood of $\mu$.  
	\end{defi}
	
	%We shall establish both upper and lower bounds of  defined above. 
	To state the results on the resolution limit to stable support recovery, we need to introduce one more concept: the super-resolution factor which is usually utilized to characterize the ill-posedness of the super-resolution problem \cite{candes2014towards}. It is defined as the ratio between Rayleigh limit and the grid scale in the grid setting and the minimum separation distance in the off-the-grid setting. In our case, since the Rayleigh limit is $\frac{\pi}{\Omega}$, we define the super-resolution factor as 
	\[SRF:= \frac{\pi}{\Omega d_{\min}},\]
	where $d_{\min}=\min_{p\neq j}|y_p-y_j|$. We have the following theorem.
	\begin{thm}\label{upperboundsupportlimithm0}
		Let $n\geq 2$, assume that $\mu=\sum_{j=1}^{n}a_j \delta_{y_j}$ is supported on $I(n, \Omega)$ and that 
		\begin{equation}\label{supportlimithm0equ0}
		\min_{p\neq j}|y_p-y_j|\geq \frac{5.88\pi e}{\Omega }\Big(\frac{\sigma}{m_{\min}}\Big)^{\frac{1}{2n-1}}. \end{equation}
		If $\hat \mu=\sum_{j=1}^{n}\hat a_{j}\delta_{\hat y_j}$ supported on $I(n,\Omega)$ is a $\sigma$-admissible measure for the measurement generated by $\mu$, then $\hat \mu$ is within the $\frac{d_{\min}}{2}$-neighborhood of $\mu$. Moreover, after reordering the $\hat y_j$'s, we have 
		\begin{equation}\label{supportlimithm0equ2}
		\Big|\hat y_j-y_j\Big|\leq \frac{C(n)}{\Omega}SRF^{2n-2}\frac{\sigma}{m_{\min}}, \quad 1\leq j\leq n,
		\end{equation}
		where $C(n)=n2^{4n-2}e^{2n}\pi^{-\frac{1}{2}}$.
	\end{thm}
	
Theorem \ref{upperboundsupportlimithm0} gives an upper bound to the computational resolution limit $\mathcal{D}_{supp}$ for the support recovery. 
%It states that when the line spectra are separated further than the upper bound in (\ref{supportlimithm0equ0}) we can stably recover their supports and any $n$-sparsity $\sigma$-admissible measure supported in $I(n, \Omega)$ preserves the structure of real one. 
%Moreover, it implies that the inverse problem of recovering the spectral supports can be regularized by the minimum separation condition with separation distance greater than the upper bound of the resolution limit in Theorem \ref{upperboundsupportlimithm0}.
%In that case, any algorithm looking for the sparsest admissible measure can achieve stable recovery. 
Compared with the Rayleigh limit $\frac{\pi}{\Omega}$, the upper bound indicates stable recovery of the supports of the line spectra in the sub-Rayleigh regime is possible if the SNR is sufficiently large.
We next show that the order of 
%the lower bound for the separation distance to ensure a stable support recovery is of the order $O(\frac{1}{\Omega}\big(\frac{\sigma}{m_{\min}}\big)^{\frac{1}{2n-1}})$, which demonstrates that our 
the upper bound is optimal.
	
	\begin{prop}\label{supportlowerboundthm0}
		For given $0<\sigma<m_{\min}$ and integer $n\geq 2$, let 
		\begin{equation}\label{supportlowerboundequ0}
		\tau=\frac{0.49e^{-\frac{3}{2}}}{\Omega}\ \Big(\frac{\sigma}{m_{\min}}\Big)^{\frac{1}{2n-1}}.
		\end{equation}
Then there exist a measure $\mu=\sum_{j=1}^{n}a_j \delta_{y_j}$ with $n$ supports at $\{-\tau, -2\tau, -n\tau\}$ and a measure $\hat \mu=\sum_{j=1}^{n}\hat a_j \delta_{\hat y_j}$ with $n$ supports at  $\{0,\tau,\cdots, (n-1)\tau\}$ such that
		$||[\hat \mu]-[\mu]||_{\infty}< \sigma$
and either $\min_{1\leq j\leq n}|a_j|= m_{\min}$ or $\min_{1\leq j\leq n}|\hat a_j|= m_{\min}$.   
	\end{prop}

Proposition \ref{supportlowerboundthm0} provides a lower bound to the computational resolution limit $\mathcal{D}_{supp}$. Combined with Theorem \ref{upperboundsupportlimithm0}, it reveals that the computational resolution limit of stable support recovery $\mathcal D_{supp}$ is of the order $O(\frac{(1/SNR)^{\frac{1}{2n-1}}}{\Omega})$. 	
	
	\begin{remark} \label{rem-diff}
We have quantitatively characterized the resolution limit to both the number detection and the support recovery problem in the LSE.
% by using the cut-off frequency, the SNR and the sparsity of the line spectra. 
The results imply that for sufficiently high SNR, the number detection problem has a better resolution limit than the support recovery problem.  
	\end{remark}
	
	\begin{remark}
		Our results imply that phase transition may occur in both the number detection and the support recovery problem in the LSE problem. See Section \ref{sec-phase} for detail. 
	\end{remark}

%   	\begin{remark}
%   		 We note that the LSE problem is closely related to the deconvolution problem in imaging where the measurement is a convolution of source and a band-limited point spread function, i.e., $\mu*f(x)=\sum_{j=1}^{n}a_j f(t-y_j)$. In the presence of additive noise $\epsilon(t)$, the measured image is 
%   	\[
%   	y(t)=\mu*f(t)+\epsilon(t)=\sum_{j=1}^{n}a_j f(t-y_j)+\epsilon(t).
%   	\]
%   	By taking the Fourier transform on both sides, we obtain 
%   	\begin{equation} \label{eq-deconv}
%   	\mathcal{F} y(x)= \mathcal{F}f(x)\cdot \mathcal F \mu(x)+ \mathcal{F} \epsilon(x)= \mathcal{F}f(x)(\sum_{j=1}^n a_j e^{i y_jx})+ \mathcal{F} \epsilon(x),
%   	\end{equation}
%   	which is reduced to our LSE problem. Assuming $0<c_1\leq \mathcal{F}f(x)\leq c_2$, according to the above formulation, it is straightforward to derive similar upper and lower bounds for the computational resolution limits of both number detection and support recovery in the deconvolution problem. 
%   \end{remark}

	\section{Approximation theory in the Vandermonde space}\label{section:approxinvande}
We present the main technique that is used in the proofs of the main results in the previous section, the approximation theory in the Vandermonde space in this section. The 
theory was first introduced in \cite{liu2019computational} and was restricted to the case of real vectors. We shall extend the theory to complex vectors. 
%Since most arguments are similar, we shall only highlight the main differences and refer the readers to \cite{liu2019computational} for more detail. 
Specifically,  for integer $s$ and $z\in \mathbb C$, we define the complex Vandermonde-vector
	\begin{equation}\label{phiformula}
	\phi_s(z)=(1,z,\cdots,z^s)^T.
	\end{equation}
%	We also define the Vandermonde space as 
%	\begin{equation*}
%	\begin{array}{l}
%	W_{s}=\spn \Big\{\ \phi_s(\omega):\omega \in\mathbb C \ \Big\}, \\ 
%	\end{array}
%	\end{equation*}
%	and $k$ dimensional Vandermonde subspace as
%	\[W_s^{k}(\omega_1,\cdots,\omega_k):=\spn \Big\{\ \phi_s(\omega_1), \ \cdots,\ \phi_s(\omega_k)\ \Big\}.\]
We consider the following non-linear least square problem:  %in the space $\spn \Big\{\ \phi_s(\omega_1), \ \cdots,\ \phi_s(\omega_k)\ \Big\}$: 
\begin{equation}\label{non-linearapproxproblem1}
	\min_{\hat a_j\in \mathbb{C},\hat \theta_j\in \mathbb{R},j=1,\cdots,k}\Big|\Big|\sum_{j=1}^k \hat a_j\phi_s(e^{i\hat \theta_j})-v\Big|\Big|_2,
\end{equation}
where $v=\sum_{j=1}^{k+1}a_j\phi_s(e^{i\theta_j})$ is given with $\theta_j$'s being real numbers. 
%This is a non-linear least square problem. 
We shall derive a lower bound for the optimal value of the minimization problem for the case when $s=k$ which is relevant to our LSE problem. The main results are presented in Section \ref{sec-vander2}.

\subsection{Notation and Preliminaries}
We introduce some notations and technical lemmas that are used in the proofs in Section \ref{sec-vander2}. 
We denote for integer $k\geq 1$, 
\begin{equation}\label{zetaxiformula1}
	\zeta(k)= \left\{
	\begin{array}{cc}
	(\frac{k-1}{2}!)^2,& \text{$k$ is odd,}\\
	(\frac{k}{2})!(\frac{k-2}{2})!,& \text{$k$ is even,}
	\end{array} 
	\right.	  \quad \xi(k)=\left\{
	\begin{array}{cc}
	\frac{1}{2},  & k=1,\\
	\frac{(\frac{k-1}{2})!(\frac{k-3}{2})!}{4},& \text{$k$ is odd,\,\,$ k\geq 3$,}\\
	\frac{(\frac{k-2}{2}!)^2}{4},& \text{$k$ is even}.
	\end{array} 
	\right.	
\end{equation}
We also define for postive integers $p, q$, and $z_1, \cdots, z_p, \hat z_1, \cdots, \hat z_q \in \mathbb C$, the following vector in $\mathbb{R}^p$
	\begin{equation}\label{notation:eta}
	\eta_{p,q}(z_1,\cdots,z_{p}, \hat z_1,\cdots,\hat z_q)=\left(\begin{array}{c}
	|(z_1-\hat z_1)|\cdots|(z_1-\hat z_q)|\\
	|(z_2-\hat z_1)|\cdots|(z_2-\hat z_q)|\\
	\vdots\\
	|(z_{p}-\hat z_1)|\cdots|(z_{p}-\hat  z_q)|
	\end{array}\right).
	\end{equation}
For complex matrix $A$, we denote $A^*$ its conjugate transpose. 
%A complex vector is viewed as a matrix. 
	
	\begin{lem}\label{projectvolumeratiolem1}
		For $s\times k$ matrix $A$ of rank $k$ with $s>k$, let $V$ be the space spanned by columns of $A$ and $V^{\perp}$ be the orthogonal complement of $V$. Denote $P_{V^{\perp}}$ the orthogonal projection to $V^{\perp}$, and $D=(A,v)$. We have
\[
		\min_{a\in \mathbb C^{k}}||Aa-v||_2=||P_{V^{\perp}}(v)||_2=	
		\sqrt{\frac{\det(D^*D)}{\det(A^*A)}}.
\]
\end{lem}
	Proof: Since $D=(A,v)$, we have 
	\begin{equation*}
	D^*D=\left(
	\begin{array}{cc}
	A^*A & A^*v\\
	v^*A & v^*v
	\end{array}
	\right).
	\end{equation*}
	By column transform we have 
	\begin{equation*}
	\det(D^*D)=\det(A^*A)\det(v^*v-v^*A(A^*A)^{-1}A^*v).
	\end{equation*}
     We decompose $v=v_1+v_2$ where $v_1\in V$ and $v_2\in V^\perp$. Then	\begin{align*}
	&\sqrt{\frac{\det(D^*D)}{\det(A^*A)}}=\sqrt{\det(v^*v-v^*A(A^*A)^{-1}A^*v)}\\
	=&\sqrt{\det(v_1^*v_1+v_2^*v_2-v_1^*A(A^*A)^{-1}A^*v_1)}.
	\end{align*}
	Note that $A(A^*A)^{-1}A^*$ is the orthogonal projection onto the space $V$. Therefore 
	$$v_1^*A(A^*A)^{-1}A^*v_1=v_1^*v_1.$$
	It follows that
	\begin{align*}
	\sqrt{\frac{\det(D^*D)}{\det(A^*A)}} =\sqrt{v_2^*v_2} =\min_{a\in \mathbb C^{k}}||Aa-v||_2.
	\end{align*}
	This completes the proof.
	
%\medskip
%	We denote for positive integers $s$ and $k$, 
%	\begin{equation}\label{vandermondematrix1}
%	V_{k}(s)=\big(\phi_{s}(e^{i \theta_1}),\cdots,\phi_s(e^{i\theta_k})\big), \quad 
%	V_{k}(k-1)=\big(\phi_{k-1}(e^{i \theta_1}),\cdots,\phi_{k-1}(e^{i\theta_k})\big).
%	\end{equation}
%	where $\phi_s(\omega)$ is the Vandermonde-vector defined in (\ref{phiformula}).
%	We now present several useful lemmas. 
	
	\medskip
\begin{lem}\label{vandemondevolumeratio1}
Let $\theta_j\in\mathbb R, j=1,\cdots,k$, and let 
$V_{k}(k)=\big(\phi_{k}(e^{i \theta_1}),\cdots,\phi_k(e^{i\theta_k})\big), 
V_{k}(k-1)=\big(\phi_{k-1}(e^{i \theta_1}),\cdots,\phi_{k-1}(e^{i\theta_k})\big)$, where $\phi_k(z)$ is defined in (\ref{phiformula}). We have
		\begin{equation*}
		\sqrt{\frac{\det\big(V_{k}(k)^*V_k(k)\big)}{\det\big(V_{k}(k-1)^*V_k(k-1)\big)}}\leq 2^{k}.
		\end{equation*}
\end{lem}
	Proof: The calculation of $\det\big(V_{k}(k-1)^*V_k(k-1)\big)$ is straightforward since $V_{k}(k-1)$ is a square Vandermonde matrix. 
The calculation of $\det\big(V_{k}(k)^*V_k(k)\big)$ is more technical. It is based on a reduced form of $V_{k}(k)$ which is obtained by applying Gaussian elimination to $V_{k}(k)$ using elementary column transformations. See Lemma 3.1, 3.2 in \cite{liu2019computational} for detail. Here we also used the fact that $|e^{i\theta_j}|=1$. 
	
	\begin{lem}\label{norminversevandermonde2}
		Let $k \geq 2$ and $-\frac{\pi}{2}\leq \theta_1<\theta_2<\cdots<\theta_{k}  \leq \frac{\pi}{2}$. Let
$\theta_{\min}=\min_{p\neq j}|\theta_p-\theta_j|$ and $V_{k}(k-1)=\big(\phi_{k-1}(e^{i \theta_1}),\cdots,\phi_{k-1}(e^{i\theta_k})\big)$. Then
		\[
		||V_k(k-1)^{-1}||_{\infty}\leq \frac{\pi^{k-1}}{\zeta(k) \theta_{\min}^{k-1}}. 
		\] 
		%where $\zeta(k)$ is defined in (\ref{zetaxiformula1}).
	\end{lem}
	Proof: Using the estimate of $\infty$-norm of the inverse of Vandermonde matrix, see for instance Theorem 1 in \cite{gautschi1962inverses}, we have
\begin{equation}\label{equ:invervande1}
	||V_k(k-1)^{-1}||_{\infty}\leq \max_{1\leq j\leq k}\Pi_{1\leq p\leq k,p\neq j}\frac{1+|e^{i\theta_p}|}{|e^{i\theta_j}-e^{i\theta_p}|} = \max_{1\leq j\leq k}\Pi_{1\leq p\leq k,p\neq j}\frac{2}{|e^{i\theta_j}-e^{i\theta_p}|}. 
\end{equation}
On the other hand, note that 
\begin{equation} \label{eq-theta}
|e^{i\theta_j}-e^{i\theta_p} | \geq \frac{2}{\pi} |\theta_j-\theta_p|, \quad \mbox{for all}\,\, \theta_j, \theta_p \in \Big[-\frac{\pi}{2}, \frac{\pi}{2}\Big],
\end{equation}
%$|e^{i\theta_j}-e^{i\theta_p}|\geq \frac{2}{\pi}|\theta_j-\theta_{p}|$, 
we have
	\begin{align*}
	&\Pi_{1\leq p\leq k,p\neq j}\frac{1}{|e^{i\theta_j}-e^{i\theta_p}|}\
	\leq  (\frac{\pi}{2})^{k-1} \Pi_{1\leq p\leq k,p\neq j}\frac{1}{|\theta_j-\theta_p|} \\
	=&(\frac{\pi}{2})^{k-1}\Pi_{p<j}\frac{1}{|\theta_j-\theta_p|} \Pi_{p>j}\frac{1}{|\theta_j-\theta_p|}
	\leq(\frac{\pi}{2})^{k-1} \frac{1}{(j-1)!\theta_{\min}^{j-1}}\ \frac{1}{(k-j)!\theta_{\min}^{k-j}} \\	
	=&\frac{1}{(j-1)!(k-j)!}(\frac{\pi}{2\theta_{\min}})^{k-1}\leq\frac{1}{\zeta(k)}(\frac{\pi}{2\theta_{\min}})^{k-1}.
	\end{align*}
The desired estimate follows. 
	
	%The next two lemmas on complex Vandermonde matrices can be proved using the same arguments as for the case of real Vandermonde matrices in \cite{liu2019computational}. 
	\begin{lem}\label{singularvaluevandermonde2}
Let $V_{k}(k-1)=\big(\phi_{k-1}(e^{i \theta_1}),\cdots,\phi_{k-1}(e^{i\theta_k})\big)$ and $V_{k}(s)=\big(\phi_{s}(e^{i \theta_1}),\cdots,\phi_s(e^{i\theta_k})\big)$
with $s>k-1$, then the following estimate on their singular values hold:
		\[\frac{1}{\sqrt{k}}\min_{1\leq j\leq k}\Pi_{1\leq p\leq k,p\neq j}\frac{|e^{i\theta_j}-e^{i\theta_p}|}{2} \leq\frac{1}{||V_k(k-1)^{-1}||_{2}}\leq  \sigma_{\min}\big(V_k(k-1)\big)\leq \sigma_{\min}\big(V_k(s)\big).\]
	\end{lem}
	Proof: The first inequality follows from (\ref{equ:invervande1}) and the fact that $||V_k(k-1)^{-1}||_{2}\leq \sqrt{k}||V_k(k-1)^{-1}||_{\infty}$. The last two inequalities are straightforward to check.  See also Lemma 9.2 in \cite{liu2019computational} or Proposition 4.4 in \cite{batenkov2020conditioning}.

 \begin{lem}{\label{lem:invervandermonde}}
Let $t_1, \cdots, t_k$ be $k$ different real numbers and let $t$ be a real number. We have
    	\[
    	\left(D_k(k-1)^{-1}\phi_{k-1}(t)\right)_{j}=\Pi_{1\leq q\leq k,q\neq j}\frac{t- t_q}{t_j- t_q},
        \]
    	where $D_k(k-1):=  \big(\phi_{k-1}(t_1),\cdots,\phi_{k-1}(t_k)\big)$. 
	    \end{lem}
Proof: 	
 We denote $\left(D_{k}(k-1)^{-1}\right)_{jq}=b_{jq}$. Observe that
 $$
 \left(D_k(k-1)^{-1}\phi_{k-1}(t)\right)_{j}=\sum_{q=1}^{k}b_{jq}t^{q-1}.
 $$
We have 
	\[\sum_{q=1}^{k}b_{jq}(t_p)^{q-1}=\delta_{jp},\ \forall j,p=1,\cdots,k,\]
where $\delta_{jp}$ is the Kronecker delta function. Then the polynomial $P_j(x)=\sum_{q=1}^{k}b_{jq}x^{q-1}$ satisfies $P_{j}(t_1)=0,\cdots,P_j(t_{j-1})=0,P_j(t_j)=1,P_j(t_{j+1})=0,\cdots,P_j(t_k)=0$. Therefore, it must be the Lagrange polynomial
	\[
	P_j(x)=\Pi_{1\leq q\leq k,q\neq j}\frac{x-t_q}{t_j-t_q}.
	\]
It follows that
	\begin{align*}
	\left(D_k(k-1)^{-1}\phi_{k-1}(t)\right)_{j}=\Pi_{1\leq q\leq k,q\neq j}\frac{ t- t_q}{t_j-t_q}.
	\end{align*}

	\begin{lem}\label{lem:multiproductlowerbound0}
Let $-\frac{\pi}{2}\leq \theta_1<\theta_2<\cdots<\theta_{k+1}  \leq \frac{\pi}{2}$ and let $\theta_{\min}=\min_{p\neq j}|\theta_p-\theta_j|$. Then 
		\[\min_{\hat \theta_1\in \mathbb R, \cdots,\hat \theta_k\in \mathbb R}||\eta_{k+1,k}(\theta_1,\cdots,\theta_{k+1},\hat \theta_1,\cdots,\hat \theta_k)||_{\infty}\geq \xi(k)(\theta_{\min})^k,\]
		where  $\xi(k)$ is defined in (\ref{zetaxiformula1}) and $\eta_{k+1,k}$ defined in (\ref{notation:eta}).
	\end{lem}
	Proof: We first show that we can find a minimizer  $(\hat \theta_1, \hat \theta_2, \cdots, \hat \theta_k)$ such that $\hat \theta_j \in [\theta_j, \theta_{j+1}), 1\leq j \leq k$. It follows that $\theta_1\leq \hat \theta_1 < \theta_2\leq \hat \theta_2<\cdots <\theta_{k+1}$. Then $||\eta_{k+1,k}||_{\infty}$ can be estimated directly.  
%	by considering $\hat \theta_j -\theta_j$ is less than $\frac{\theta_{\min}}{2}$ or not. 
We refer to Lemma 3.4 in \cite{liu2019computational} for the detail. 
	
	\medskip
Using (\ref{eq-theta}),
%\begin{equation} \label{eq-theta}
%|e^{i\theta_j}-e^{i\theta_p} | \geq \frac{2}{\pi} |\theta_j-\theta_p|, \quad \mbox{for all}\,\, \theta_j, \theta_p \in \Big[-\frac{\pi}{2}, \frac{\pi}{2}\Big],
%\end{equation}
we can extend Lemma \ref{lem:multiproductlowerbound0} to the complex case. 
	
\begin{cor}\label{cor:multiproductlowerbound1}
Let $-\frac{\pi}{2}\leq \theta_1<\theta_2<\cdots<\theta_{k+1}  \leq \frac{\pi}{2}$. Assume that $\min_{p\neq j}|\theta_p-\theta_j|=\theta_{\min}$, then for any $\hat \theta_1,\cdots, \hat \theta_k\in \mathbb R$, we have the following estimate 	
		\[||\eta_{k+1,k}(e^{i\theta_1},\cdots,e^{i\theta_{k+1}},e^{i \hat \theta_1},\cdots,e^{i\hat \theta_k})||_{\infty}\geq \xi(k)(\frac{2 \theta_{\min}}{\pi})^k.  \] 
	\end{cor}
	
	\begin{lem}\label{lem:stablemultiproduct0}
Let $-\frac{\pi}{2}\leq \theta_1<\theta_2<\cdots<\theta_k \leq \frac{\pi}{2}$ and $ \hat \theta_1, \hat \theta_2, \cdots, \hat \theta_k \in \big[-\frac{\pi}{2}, \frac{\pi}{2}\big]$. Assume that
		\begin{equation}\label{equ:satblemultiproductlemma1equ1}
		||\eta_{k,k}(\theta_1,\cdots,\theta_k, \hat \theta_1,\cdots, \hat \theta_k)||_{\infty}< \epsilon,
		\end{equation}
		where $\eta_{k,k}$ is defined as in (\ref{notation:eta}), and that
		\begin{equation}\label{equ:satblemultiproductlemma1equ2}
		\theta_{\min} =\min_{q\neq j}|\theta_q-\theta_j|\geq \Big(\frac{4\epsilon}{\lambda(k)}\Big)^{\frac{1}{k}},
		\end{equation} 
		where 
		\begin{equation*}\label{equ:lambda1}
	    \lambda(k)=\left\{
		\begin{array}{ll}
		1,  & k=2,\\
		\xi(k-2),& k\geq 3.
		\end{array} 
		\right.	
%\xi(k-2)=\left\{
%		\begin{array}{ll}
%		\frac{1}{2},  & k=3,\\
%		\frac{(\frac{k-3}{2})!(\frac{k-5}{2})!}{4},& \text{$ k\geq 5$ odd,}\\
%		\frac{(\frac{k-4}{2}!)^2}{4},& \text{$k\geq 4$ even}.
%		\end{array} 
%		\right.	
		\end{equation*}
		Then after reordering $\hat \theta_j$'s, we have
		\begin{equation}\label{equ:satblemultiproductlemma1equ4}
		|\hat \theta_j -\theta_j|< \frac{\theta_{\min}}{2},  \quad j=1,\cdots,k,
		\end{equation}
		and moreover
		\begin{equation}\label{equ:satblemultiproductlemma1equ5}
		|\hat \theta_j -\theta_j|\leq \frac{2^{k-1}\epsilon}{(k-2)!(\theta_{\min})^{k-1}}, \quad j=1,\cdots, k.
		\end{equation}
	\end{lem}

Proof: 
\textbf{Step 0.} We only prove the lemma for $k\geq 3$ and the case $k=2$ can be deduced in a similar manner. 

\textbf{Step 1.} We claim that 
for each $\hat \theta_p, 1\leq p\leq k$, there exists one $\theta_j$ such that $|\hat \theta_p-\theta_j|<\frac{\theta_{\min}}{2}$.
By contradiction, suppose there exists $p_0$ such that $|\theta_j - \hat \theta_{p_0}|\geq \frac{\theta_{\min}}{2}$ for all $1\leq j\leq k$. 
%We exclude the specific $\hat \theta_{p_0}$ term from $\eta_{k,k}$ and decompose 
Observe that 
$$
\eta_{k,k}(\theta_1,\cdots,\theta_k, \hat \theta_1,\cdots, \hat \theta_k)
=\text{diag}\left((\theta_1- \hat \theta_{p_0}),\cdots,(\theta_{k}-\hat \theta_{p_0})\right)\eta_{k,k-1}(\theta_1,\cdots,\theta_{k}, \hat \theta_1,\cdots,\hat \theta_{p_0-1},\hat \theta_{p_0+1},\cdots,\hat \theta_k).
$$ 
Using Lemma \ref{lem:multiproductlowerbound0}, we have
\[
	||\eta_{k,k}||_{\infty}\geq \frac{\theta_{\min}}{2}||\eta_{k, k-1}||_{\infty} \geq  \frac{\xi(k-1)}{2}(\theta_{\min})^{k}.  
\]	
By the formula of $\xi(k)$ in (\ref{zetaxiformula1}), we can verify directly that $\frac{\xi(k-1)}{2}\geq \frac{\xi(k-2)}{4}$. 
Therefore, 
\[
||\eta_{k,k}||_{\infty}\geq \frac{\xi(k-2)}{4} (\theta_{\min})^{k} 
\geq\epsilon,
\]
where we used (\ref{equ:satblemultiproductlemma1equ2}) in the last inequality above. This contradicts to (\ref{equ:satblemultiproductlemma1equ1}) and hence proves our claim.
	
\textbf{Step 2.} We claim that for each $\theta_j, 1\leq j\leq k$, there exists one and only one $\hat \theta_p$ such that $|\theta_j -\hat \theta_p|< \frac{\theta_{\min}}{2}$.
%	By the result in Step 1, we see that all the $\hat \theta_p$ is in a $\frac{\theta_{\min}}{2}$-neighborhood of some $\theta_j$. To demonstrate 	\textbf{Claim 2}, 
It suffices to show that for each $\theta_j, 1\leq j\leq k$, there is only one $\hat \theta_p$ such that $|\theta_j-\hat \theta_p|<\frac{\theta_{\min}}{2}$. By contradiction, suppose there exist $p_1,p_2,$ and $j_0$ such that $|\theta_{j_0}-\hat \theta_{p_1}|<\frac{\theta_{\min}}{2}, |\theta_{j_0}-\hat \theta_{p_2}|<\frac{\theta_{\min}}{2}$. Then for all $j \neq j_0$, we have 
	\begin{equation}\label{equ:satblemultiproductlemma1equ3}
	|(\theta_j- \hat \theta_{p_1}) (\theta_j-\hat \theta_{p_2})|\geq \frac{(\theta_{\min})^2}{4}.
	\end{equation}
Similar to the argument in Step 1, we separate the factors involving $\hat \theta_{p_1}, \hat \theta_{p_2}, \theta_{j_0}$ from $\eta_{k,k}$ and  consider
\[
\eta_{k-1,k-2}=	\eta_{k-1,k-2}(\theta_1,\cdots,\theta_{j_0-1},\theta_{j_0+1},\cdots,\theta_k, \hat \theta_1,\cdots,\hat \theta_{p_1-1},\hat \theta_{p_1+1}, \cdots, \hat \theta_{p_2-1}, \hat \theta_{p_2+1},\cdots, \hat \theta_k).
\]
Note that the components of $\eta_{k-1,k-2}$ differ from those of $\eta_{k,k}$ only by the factors $|(\theta_j-\hat \theta_{p_1})(\theta_{j}-\hat \theta_{p_2})|$ for $j=1,\cdots,j_0-1,j_0+1,\cdots,k$. We can show that
$$	
||\eta_{k,k}||_{\infty}\geq \frac{\theta_{\min}^2}{4} ||\eta_{k-1,k-2}||_{\infty}. 
$$
Using Lemma \ref{lem:multiproductlowerbound0} and (\ref{equ:satblemultiproductlemma1equ2}), we further get 
\begin{align*}
	||\eta_{k,k}||_{\infty}\geq \frac{\xi(k-2)}{4}(\theta_{\min})^{k}\geq \epsilon, 
\end{align*}
which contradicts to (\ref{equ:satblemultiproductlemma1equ1}). This contradiction proves our claim. 
	
\textbf{Step 3.} By the result in Step 2,  we can reordering $\hat \theta_j$'s to get
	\[
	|\hat \theta_j -\theta_j|< \frac{\theta_{\min}}{2}, \quad j=1,\cdots,k.
	\]

\textbf{Step 4.} We prove (\ref{equ:satblemultiproductlemma1equ5}). By (\ref{equ:satblemultiproductlemma1equ4}), it is clear that
	\begin{equation}\label{equ:satblemultiproductlemma1equ6}
	|\theta_{j} - \hat \theta_{p}|\geq \left\{
	\begin{array}{cc}
	(j-p-\frac{1}{2})\theta_{\min} & p<j, \\
	(p-j-\frac{1}{2})\theta_{\min} & p>j.
	\end{array}
	\right.
	\end{equation} 
	We next show that 
	\begin{equation} \label{eq-222}
	|(\theta_j-\hat \theta_1)\cdots(\theta_j-\hat \theta_k)|\geq |\theta_{j}-\hat \theta_j|(\frac{\theta_{\min}}{2})^{k-1} (k-2)!, \quad j=1, 2, \cdots, k. 
	\end{equation}  
	Indeed, for $2\leq j\leq k-1$, we have
	\begin{align*}
	&|(\theta_j-\hat \theta_1)\cdots(\theta_j-\hat \theta_k)|=|(\theta_j-\hat \theta_j)| \Pi_{1\leq p\leq j-1}|(\theta_j-\hat \theta_p)|\Pi_{j+1\leq p\leq k}|(\theta_j-\hat \theta_p)|\\
	\geq &|(\theta_j-\hat \theta_j)|\ \Big(\Pi_{1\leq p\leq j-1}\frac{2(j-p)-1}{2}\theta_{\min}\Big) \Big(\Pi_{j+1\leq p\leq k} \frac{2(p-j)-1}{2}\theta_{\min}\Big) \quad \Big(\text{by (\ref{equ:satblemultiproductlemma1equ6})}\Big)\\
	=& |\theta_{j}-\hat \theta_j|(\frac{\theta_{\min}}{2})^{k-1}(2j-3)!!(2(k-j)-1)!!\quad  \,\, \\
	\geq& |\theta_{j}-\hat \theta_j|(\frac{\theta_{\min}}{2})^{k-1} (k-2)!. \quad \quad \Big(\mbox{since $(2j-3)!!(2(k-j)-1)!!\geq (k-2)!$}\Big)
	\end{align*}
	Similarly, we can prove (\ref{eq-222}) for $j=1$ and $j=k$. Combining  (\ref{eq-222}) and (\ref{equ:satblemultiproductlemma1equ1}), we get 
	\[
	|\hat \theta_{j}- \theta_j|(\frac{\theta_{\min}}{2})^{k-1}(k-2)!< \epsilon, \quad j=1, 2, \cdots, k,
	\]
whence (\ref{equ:satblemultiproductlemma1equ5}) follows. This completes the proof of the lemma.

\medskip

By (\ref{eq-theta}), Lemma \ref{lem:stablemultiproduct0} leads to the following corollary.
\begin{cor}\label{cor:stablemultiproduct1}
	For $-\frac{\pi}{2}\leq \theta_1<\theta_2<\cdots<\theta_k \leq \frac{\pi}{2}$ and $ \hat \theta_1, \hat \theta_2, \cdots, \hat \theta_k \in [-\frac{\pi}{2}, \frac{\pi}{2}]$, if
	\begin{equation*}
	||\eta_{k,k}(e^{i \theta_1},\cdots,e^{i \theta_k}, e^{i \hat \theta_1},\cdots, e^{i \hat \theta_k})||_{\infty}< (\frac{2}{\pi})^{k}\epsilon, \text{ and } \theta_{\min} =\min_{q\neq j}|\theta_q-\theta_j|\geq  \Big(\frac{4\epsilon}{\lambda(k)}\Big)^{\frac{1}{k}},
	\end{equation*}
	where $\eta_{k,k}$ defined by (\ref{notation:eta}) and $\lambda(k)$ is defined as in (\ref{equ:lambda1}),
	then after reordering $\hat \theta_j$'s, we have
	\begin{equation}\label{equ:satblemultiproductcor1}
	|\hat \theta_j -\theta_j|< \frac{\theta_{\min}}{2}  \text{ and }
	|\hat \theta_j -\theta_j|\leq \frac{2^{k-1}\epsilon}{(k-2)!(\theta_{\min})^{k-1}}, \quad j=1,\cdots, k.
	\end{equation}
\end{cor}

	\subsection{Main results on the approximation theory in the Vandermonde space} \label{sec-vander2}
	We first derive a lower bound for the non-linear approximation problem (\ref{non-linearapproxproblem1}) when $v=\phi_{k}(e^{i\hat \theta})$ is a  Vandermonde-vector. 
%	It is an important theorem that estimates the approximation in Vandermonde space by the distance between matrix elements, e.g., $|e^{i\theta}-e^{i\hat \theta_j}|$ in the following theorem. 
	
	\begin{thm}\label{spaceapprolowerbound0}
		Let $k\geq 1$, for fixed $\hat \theta_1,\cdots, \hat \theta_k\in \mathbb{R}$, denote $\hat A= \big(\phi_{k}(e^{i\hat \theta_1}),\cdots, \phi_{k}(e^{i\hat \theta_k})\big)$ where  $\phi_{k}(e^{i\hat \theta_j})$'s  are defined as in (\ref{phiformula}). Let $V$ be the $k$ dimensional complex space spanned by the column vectors of $\hat A$, and $V^\perp$ the one dimensional orthogonal complement of $V$ in $\mathbb{C}^{k+1}$. Let $P_{V^{\perp}}$ be the orthogonal projection onto $V^{\perp}$ in $\mathbb{C}^{k+1}$, we have
		\begin{equation*}
		\min_{\hat a\in \mathbb C^{k}}||\hat A\hat a-\phi_{k}(e^{i\theta})||_2=||P_{V^{\perp}}\big(\phi_{k}(e^{i\theta})\big)||_2 = |v^*\phi_{k}(e^{i\theta}) |\geq \frac{1}{2^k}|\Pi_{j=1}^k(e^{i\theta}-e^{i\hat \theta_j})|.
		\end{equation*}	
where $v$ is a unit vector in $V^\perp$ and $v^*$ is its conjugate transpose. 	
	\end{thm}
Proof: By Lemma \ref{projectvolumeratiolem1},
\[
	\min_{\hat a\in \mathbb C^{k}}||\hat A\hat a-\phi_{k}(e^{i\theta})||_2 = \sqrt{\frac{\det(D^*D)}{\det(\hat A^*\hat A)}},
\]
where $D=\big(\phi_{k}(e^{i\hat \theta_1}),\cdots, \phi_{k}(e^{i\hat \theta_k}), \phi_{k}(e^{i\theta})\big)$. 
Denote $\tilde{A}= \big(\phi_{k-1}(e^{i\hat \theta_1}),\cdots, \phi_{k-1}(e^{i\hat \theta_k})\big)$. 
By Lemma \ref{vandemondevolumeratio1}, we have
$$
\sqrt{\frac{\det(\hat A^*\hat A)}{\det(\tilde A^*\tilde A)}} \leq 2^k. 
$$
Therefore
\[
\min_{\hat a\in \mathbb C^{k}}||\hat A\hat a-\phi_{k}(e^{i\theta})||_2 \geq \frac{1}{2^k}\sqrt{\frac{\det(D^*D)}{\det(\tilde A^*\tilde A)}}.
\]
Note that $D$ and $\tilde{A}$ are square Vandermonde matrices. We can use the formula for their determinant to derive that
\[
\min_{\hat a\in \mathbb C^{k}}||\hat A\hat a-\phi_{k}(e^{i\theta})||_2 \geq \frac{1}{2^k}\frac{|\Pi_{1\leq t<p \leq k}(e^{i \hat \theta_t}-e^{i \hat \theta_p})\Pi_{q=1}^{k}(e^{i \theta}-e^{i \hat \theta_q})|}{|\Pi_{1\leq t<p \leq k}(e^{i \hat \theta_t}-e^{i \hat \theta_p})|}= \frac{1}{2^k}|\Pi_{j=1}^k(e^{i\theta}-e^{i\hat \theta_j})|.
\]
The remaining statements of the theorem is straightforward to show. This completes the proof of the theorem.

\medskip
We then show a sharp lower bound for the non-linear approximation problem (\ref{non-linearapproxproblem1}) when $v=\sum_{j=1}^{k+1}a_j\phi_s(e^{i\theta_j})$ is the linear combination of $k+1$ Vandermonde vectors. 
\begin{thm}\label{spaceapprolowerbound1}
Let $k\geq 1$. Assume $\theta_j\in\big[\frac{-\pi}{2}, \frac{\pi}{2}\big], 1\leq j\leq k+1$ are $k+1$ distinct points, and $|a_j|\geq m_{\min},1\leq j\leq k+1$. Let $\theta_{\min}=\min_{p\neq j}|\theta_p-\theta_j|$. For $q\leq k$, let $\hat a(q)=(\hat a_1,\cdots, \hat a_q)^T$, $a=(a_1,\cdots, a_{k+1})^T$ and
\[\hat A(q)= \big(\phi_{2k}(e^{i\hat \theta_1}),\cdots, \phi_{2k}(e^{i\hat \theta_q})\big), \quad A= \big(\phi_{2k}(e^{i\theta_1}),\cdots, \phi_{2k}(e^{i\theta_{k+1}})\big)\]
where $\phi_{2k}(z)$ is defined as in (\ref{phiformula}). Then
\begin{equation*}
\min_{\hat a_p\in \mathbb C,\hat \theta_p\in \mathbb R,p=1,\cdots,q}||\hat A(q)\hat a(q)-Aa||_2\geq  \frac{\zeta(k+1)\xi(k)m_{\min}\theta_{\min}^{2k}}{\pi^{2k}}.
\end{equation*}
\end{thm}
Proof: \textbf{Step 1.} Note that for $q<k$, we have
\[
  	\min_{\hat a_p\in \mathbb C,\hat \theta_p\in \mathbb R,p=1,\cdots,q}||\hat A(q)\hat a(q)-Aa||_2\geq 	\min_{\hat a_p\in \mathbb C,\hat \theta_p\in \mathbb R,p=1,\cdots,k}||\hat A(k)\hat a(k)-Aa||_2.
\]
So we only need to prove the case when $q=k$. In addition, it suffices to show that for any $\hat \theta_1,\cdots,\hat \theta_{k}\in \mathbb R, 1\leq p\leq k$, we have 
\begin{equation}\label{spaceapproxlowerboundequ-3}
\min_{\hat a\in \mathbb{C}^k}||\hat A(k)\hat a(k)-Aa||_2\geq \frac{\zeta(k+1)\xi(k)m_{\min}(\theta_{\min})^{2k}}{\pi^{2k}}.
\end{equation}
So we fix $(\hat \theta_1,\cdots,\hat \theta_{k})$ in our subsequent argument.\\
\textbf{Step 2.}  
%The proof of the theorem is essentially based on a key trick that we traverse the matrices $\hat A(k)$ and $A$, and choose circularly the partial matrices $\hat A_{l}$ and $A_l$ from them that 
For $l=0, \cdots, k$, we define the partial matrices
\[
\hat A_{l}=
\begin{pmatrix}
e^{il \hat \theta_1}&\cdots&e^{il \hat \theta_k}\\
e^{i(l+1) \hat \theta_1}&\cdots &e^{i(l+1)\hat \theta_k}\\
\vdots &\vdots &\vdots\\ 
e^{i(l+k)\hat \theta_1}&\cdots&e^{i(l+k) \hat \theta_k}
\end{pmatrix},
\quad 
A_{l}=
\begin{pmatrix}
e^{il \theta_1}&\cdots&e^{il\theta_{k+1}}\\
e^{i(l+1) \theta_1}&\cdots &e^{i(l+1)\theta_{k+1}}\\
\vdots &\vdots &\vdots\\ 
e^{i(l+k)\theta_1}&\cdots&e^{i(l+k) \theta_{k+1}}
\end{pmatrix}.
\]
%where $l=0, \cdots, k$. Each $\hat A_{l}$ and $A_l$ is a section of $\hat A(k)$ and $A$ respectively and we estimate $\min_{\hat a\in \mathbb C^k}||\hat A_l\hat a-A_l a||_2$ each time, rather than estimating (\ref{spaceapproxlowerboundequ-3}) directly. Precisely, we claim that
%\[
%\max_{l=0,\cdots,k}\ \min_{\hat a\in \mathbb C^k}||\hat A_l\hat a-A_l a||_2\geq \frac{\zeta(k+1)\xi(k)m_{\min}(\theta_{\min})^{2k}}{\pi^{2k}}.
%\]
It is clear that for all $l$, we have 
\begin{equation}\label{spaceapproxlowerboundequ-4}
\min_{\hat a\in \mathbb C^{k}}||\hat A(k) \hat a(k)- A a||_2\geq \ \min_{\hat a\in \mathbb C^{k}}||\hat A_l\hat a-A_la||_2.
\end{equation}
%we consequently prove (\ref{spaceapproxlowerboundequ-3}) and thus demonstrate the theorem. Therefore, in next step, we shall prove the claim and complete the proof.\\
\textbf{Step 3.} For each $l$, observe that $\hat A_l = \hat A_0 \text{diag}(e^{il\hat \theta_1}, \cdots, e^{il\hat \theta_k})$, $ A_l =  A_0 \text{diag}(e^{il \theta_1}, \cdots, e^{il \theta_{k+1}})$. We have 
\begin{equation}\label{spaceapproxlowerboundequ-1}\min_{\hat a\in \mathbb C^k}||\hat A_l\hat a-A_l a||_2=\min_{\hat a\in \mathbb C^{k}}||\hat A_0\hat a-A_0\gamma_l||_2,\end{equation}
where $\hat a=(\hat a_1,\cdots,\hat a_{k})^T$ and $\gamma_l=(a_1e^{il \theta_1},\cdots,a_{k+1}e^{il\theta_{k+1}})^T$. Let $V$ be the complex space spanned by the column vectors of $\hat A_0$, $V^\perp$ be the orthogonal complement of $V$ in $\mathbb C^{k+1}$. It is clear that $V^\perp$ is a one-dimensional complex space. We let $v$ be a unit vector in $V^{\perp}$ and denote $P_{V^{\perp}}$ the orthogonal projection onto $V^{\perp}$ in $\mathbb C^{k+1}$. Note that $\|P_{V^{\perp}}u\|_{2} =  |v^* u| $ for $u \in \mathbb C^{k+1}$ where
$v^*$ is the conjugate transpose of $v$. We have
\begin{align}\label{spaceapproxlowerboundequ-2}
\min_{\hat a\in \mathbb C^k}||\hat A_0\hat a-A_0\gamma_l||_2=||P_{V^{\perp}}(A_0\gamma_l)||_2= |v^*A_0\gamma_l|= |\sum_{j=1}^{k+1}a_je^{il\theta_j}v^*\phi_{k}(e^{i\theta_j})| =|\beta_l|.
\end{align} 
%where $\phi_{k}(z)$ is defined in (\ref{phiformula})  
% We can write 
%\begin{equation}\label{spaceapproxlowerboundequ0}      
%P_{V^{\perp}}(\phi_{k}(e^{i\theta_j}))=\lambda_j||P_{V^{\perp}}(\phi_{k}(e^{i\theta_j}))||_2v, \quad j=1,\cdots,k+1,
%\end{equation}
%where $|\lambda_j|=1$. 
%Combing (\ref{spaceapproxlowerboundequ-2}) and (\ref{spaceapproxlowerboundequ0}), we get 
%\begin{equation}\label{spaceapproxlowerboundequ-5}
%\min_{\hat a\in \mathbb C^k}||\hat A_0\hat a-A_0\gamma_l||_2=|\beta_l|,
%\end{equation}
where 
\[
\beta_l= \sum_{j=1}^{k+1} a_je^{il\theta_j}v^*\phi_{k}(e^{i\theta_j}), \quad \mbox{for}\,\,\,l=0, 1,\cdots, k.
\]
\textbf{Step 4.} 
Denote $\beta=(\beta_0,\cdots,\beta_{k})^T$. We have
$
\beta= B \hat \eta,
$
where 
\[B=\left(\begin{array}{cccc}
a_1&a_2&\cdots&a_{k+1}\\
a_1e^{i\theta_1}&a_2e^{i\theta_2}&\cdots&a_{k+1}e^{i\theta_{k+1}}\\
\vdots&\vdots&\vdots&\vdots\\
a_1e^{ik\theta_1}&a_2e^{ik\theta_2}&\cdots&a_{k+1}e^{ik\theta_{k+1}}
\end{array}\right),\quad \hat \eta=
\left(\begin{array}{c}
v^*\phi_{k}(e^{i\theta_1})\\
v^*\phi_{k}(e^{i\theta_2})\\
\vdots\\
v^*\phi_{k}(e^{i\theta_{k+1}})
\end{array}\right).
\]
By Lemma \ref{norminversevandermonde2}, we have 
\begin{align*}
||\hat \eta||_{\infty}=||B^{-1}\beta||_{\infty}\leq ||B^{-1}||_{\infty}||\beta||_{\infty}
\leq \frac{\pi^{k}}{ \zeta(k+1)m_{\min}\theta_{\min}^{k}}||\beta||_{\infty}.
\end{align*}
On the other hand, by Theorem \ref{spaceapprolowerbound0}, we have
\begin{equation*}\label{equ:spaceapproxlowerbound1equ2}
||\hat \eta||_{\infty} \geq \frac{1}{2^k}||\eta_{k+1, k}(e^{i\theta_1}, \cdots, e^{i\theta_{k+1}},  e^{i\hat \theta_1}, \cdots, e^{i\hat \theta_k} )||_{\infty}.
\end{equation*}
%where 
%\[\eta=
%\left(\begin{array}{c}
%|(e^{i\theta_1}- e^{i \hat\theta_1})\cdots(e^{i\theta_1}-e^{i\hat \theta_k})|\\
%|(e^{i\theta_2}- e^{i \hat\theta_1})\cdots(e^{i\theta_2}-e^{i\hat \theta_k})|\\
%\vdots\\
%|(e^{i\theta_{k+1}}- e^{i \hat\theta_1})\cdots(e^{i\theta_{k+1}}-e^{i\hat \theta_k})|\\
%\end{array}\right).
%\] 
%Equivalently,
%\[||\beta||_{\infty}\geq \frac{\zeta(k+1)m_{\min}\theta_{\min}^{k}}{2^k\pi^{k}}||\eta||_{\infty}.\]
Combining this with Corollary \ref{cor:multiproductlowerbound1}, we get
$$
||\hat \eta||_{\infty} \geq \frac{1}{2^k}\xi(k)(\frac{2 \theta_{\min}}{\pi})^k. 
$$
It follows that 
\begin{align*}
||\beta||_{\infty}\geq \frac{\zeta(k+1)\xi(k)m_{\min}(\theta_{\min})^{2k}}{\pi^{2k}}. 
\end{align*}
Therefore, recalling (\ref{spaceapproxlowerboundequ-4})-(\ref{spaceapproxlowerboundequ-2}), we have
\[\min_{\hat a\in \mathbb C^{k}}||\hat A(k) \hat a(k)- A a||_2\geq \ \max_{0\leq l \leq k} \min_{\hat a\in \mathbb C^{k}}||\hat A_l\hat a-A_la||_2 = \ \max_{0\leq l \leq k} |\beta_l| = ||\beta||_{\infty} \geq \frac{\zeta(k+1)\xi(k)m_{\min}(\theta_{\min})^{2k}}{\pi^{2k}}.
\]
This proves (\ref{spaceapproxlowerboundequ-3}) and hence the theorem. 
%or equivalently, by (\ref{spaceapproxlowerboundequ1}), 
%\[\max_{l=0,\cdots,k}\ \min_{\hat \gamma_l\in \mathbb C^{k}}||\hat A_0\hat \gamma_l-A_0\gamma_l||_2\geq \frac{\zeta(k+1)\xi(k)m_{\min}(\theta_{\min})^{2k}}{\pi^{2k}}.\]
%Thus by (\ref{spaceapproxlowerboundequ-1}),
%\[\max_{l=0,\cdots,k}\ \min_{\hat a\in \mathbb C^{k}}||\hat A_l\hat a-A_la||_2\geq \frac{\zeta(k+1)\xi(k)m_{\min}(\theta_{\min})^{2k}}{\pi^{2k}},\]
%and we prove the claim.    

\medskip
\begin{thm}\label{spaceapproxlowerbound2}
	Let $k\geq 2$. Assume $\theta_1,\cdots,\theta_{k}\in \big[\frac{-\pi}{2}, \frac{\pi}{2}\big]$ are $k$ different points and $|a_j|\geq m_{\min},1\leq j\leq k$. Define $\theta_{\min}=\min_{p\neq j}|\theta_p-\theta_j|$. Assume $k$ distinct $\hat \theta_1,\cdots,\hat \theta_k\in \big[\frac{-\pi}{2}, \frac{\pi}{2}\big]$ satisfy
	\[ ||\hat A\hat a-A a||_2< \sigma, \]
	where
	$\hat a=(\hat a_1,\cdots, \hat a_k)^T$, $a=(a_1,\cdots, a_{k})^T$ and
	\[\hat A= \big(\phi_{2k-1}(e^{i \hat \theta_1}),\cdots, \phi_{2k-1}(e^{i \hat \theta_k})\big), \quad A= \big(\phi_{2k-1}(e^{i \theta_1}),\cdots, \phi_{2k-1}(e^{i \theta_{k}})\big).\]
	Then
%	Here $\phi_{2k-1}(\omega)$ is defined as in (\ref{phiformula}). Then for $\eta_{k,k}(e^{i \theta_1},\cdots,e^{i \theta_k},e^{i \hat \theta_1},\cdots,e^{i \hat \theta_k})$ defined in (\ref{notation:eta}), we have 
	\[||\eta_{k,k}(e^{i \theta_1},\cdots,e^{i \theta_k},e^{i \hat \theta_1},\cdots,e^{i \hat \theta_k})||_{\infty}<\frac{2^{k}\pi^{k-1}}{\zeta(k)\theta_{\min}^{k-1}}\frac{\sigma}{m_{\min}}.\]
\end{thm}
Proof:  
%The proof is similar to that of Theorem 3.2 in \cite{liu2019computational} %and utilize identical ideas and techniques to the proof of Theorem \ref{spaceapprolowerbound1}. 
For ease of presentation, here we only outline the main idea and omit the details. First, similar to Step 2 in proof of Theorem \ref{spaceapprolowerbound1}, we can show that $||\hat A\hat a-A a||_2< \sigma$ implies that 
	\begin{equation}\label{equ:spaceapproxlowerbound2eq1}
	\max_{l=0,\cdots,k-1}\ \min_{\hat a\in \mathbb C^k}||\hat A_l\hat a-A_l a||_2\leq ||\hat A\hat a-A a||_2< \sigma
	\end{equation}
	where 
\[
	\hat A_{l}=
	\begin{pmatrix}
	e^{il \hat \theta_1}&\cdots&e^{il \hat \theta_k}\\
	e^{i(l+1) \hat \theta_1}&\cdots &e^{i(l+1)\hat \theta_k}\\
	\vdots &\vdots &\vdots\\ 
	e^{i(l+k)\hat \theta_1}&\cdots&e^{i(l+k) \hat \theta_k}
	\end{pmatrix},
	\quad 
	A_{l}=
	\begin{pmatrix}
	e^{il \theta_1}&\cdots&e^{il\theta_{k}}\\
	e^{i(l+1) \theta_1}&\cdots &e^{i(l+1)\theta_{k}}\\
	\vdots &\vdots &\vdots\\ 
	e^{i(l+k)\theta_1}&\cdots&e^{i(l+k) \theta_{k}}
	\end{pmatrix},
\]
are partial matrices of $\hat A(k)$ and $A$ respectively. 	
Second, similar to Step 3 and 4 of the proof of Theorem \ref{spaceapprolowerbound1}, we can show that
\[
||\eta_{k,k}(e^{i \theta_1},\cdots,e^{i \theta_k},e^{i \hat \theta_1},\cdots,e^{i \hat \theta_k})||_{\infty}\leq \frac{2^{k}\pi^{k-1}}{\zeta(k)\theta_{\min}^{k-1}m_{\min}}\max_{l=0,\cdots,k-1}\ \min_{\hat a\in \mathbb C^k}||\hat A_l\hat a-A_l a||_2,
\]
whence the theorem follows.

\section{Proofs of the main results in Section \ref{section:Mainresults}} \label{section:mainresultsproof}
	\subsection{Proof of Theorem \ref{upperboundnumberlimithm0}}	
	\textbf{Step 1.} We write 
	\begin{equation}\label{Mdecomposition1}
	M=(2n-1)r+q,
	\end{equation}
	where $r, q$ are integers with $r\geq 1$ and $0\leq q< 2n-1$. 
	We denote $\theta_j = 	y_j\frac{2r\Omega}{M-1}, j=1,\cdots,n$. For $y_j\in I(n,\Omega)=[-\frac{(n-1)\pi}{2\Omega},\frac{(n-1)\pi}{2\Omega}]$, in view of (\ref{Mdecomposition1}), it is clear that
	\begin{equation}\label{equ:distributionequ1}
	\theta_j = y_j\frac{2r\Omega}{M-1}\in \Big[\frac{-\pi}{2}, \frac{\pi}{2}\Big], \quad j=1,\cdots,n.
	\end{equation}

\textbf{Step 2.} For $\hat \mu =\sum_{j=1}^{k}\hat a_j \delta_{\hat y_j}$ with $k<n$, note that
%	 the proof of Theorem \ref{upperboundnumberlimithm0} essentially depends on the distance between measurement $[\hat \mu]$ and $[\mu]$ for which we shall derive a sharp lower bound. The measurement difference 
\[
	[\hat \mu]-[\mu] = \left(\mathcal F \hat \mu(\omega_1), \mathcal F \hat \mu(\omega_2), \cdots,\mathcal F \hat \mu(\omega_M)\right)^T-\left(\mathcal F \mu(\omega_1), \mathcal F  \mu(\omega_2), \cdots,\mathcal F \mu(\omega_M)\right)^T,
\]
where $\omega_1=-\Omega, \omega_2=-\Omega+h, \cdots,\omega_M=-\Omega+(M-1)h$ and $h=\frac{2\Omega}{M-1}$. Using 
only the partial measurement at $\omega_{1+tr}=-\Omega+trh, 0\leq t\leq 2n-2$, we have 
\begin{equation*}\label{equ:upperboundnumberlimithm0equ0}
	\left(\mathcal F \hat \mu(\omega_1), \mathcal F \hat \mu(\omega_{1+r}), \cdots,\mathcal F \hat \mu(\omega_{1+(2n-2)r})\right)^T-\left(\mathcal F \mu(\omega_1), \mathcal F \mu(\omega_{1+r}), \cdots,\mathcal F \mu(\omega_{1+(2n-2)r})\right)^T =\hat B_1 \hat a- B_1 a 
\end{equation*}
%	where $x_{1+tr}=-\Omega+trh, 0\leq t\leq 2n-2$ and $r$ is from (\ref{Mdecomposition1}). Written in a matrix form, the difference (\ref{equ:upperboundnumberlimithm0equ0}) is 
where $\hat a= (\hat a_1,\cdots, \hat a_k)^T$, $a=(a_1, \cdots, a_n)^T$ and  
\begin{equation*}
	\hat B_1= \left(
	\begin{array}{ccc}
	e^{i\hat y_1\omega_1}&\cdots& e^{i\hat y_k\omega_1}\\
	e^{i\hat y_1\omega_{1+r}} &\cdots& e^{i\hat y_k\omega_{1+r}}\\
	\vdots&\vdots&\vdots\\
	%e^{i\hat y_1\omega_{1+(2n-3)r}}&\cdots& e^{i\hat y_k\omega_{1+(2n-3)r}}\\
	e^{i\hat y_1\omega_{1+(2n-2)r}}&\cdots& e^{i\hat y_k \omega_{1+(2n-2)r}}\\
	\end{array}
	\right), \quad B_1=\left(
	\begin{array}{ccc}
	e^{i y_1\omega_1}&\cdots& e^{i y_n\omega_1}\\
	e^{i y_1\omega_{1+r}} &\cdots& e^{i y_n\omega_{1+r}}\\
	\vdots&\vdots&\vdots\\
	%e^{i y_1\omega_{1+(2n-3)r}}&\cdots& e^{i y_n\omega_{1+(2n-3)r}}\\
	e^{i y_1\omega_{1+(2n-2)r}}&\cdots& e^{i y_n \omega_{1+(2n-2)r}}\\
	\end{array}
	\right).
\end{equation*}
It is clear that 
	\begin{equation} \label{equ:upperboundnumberlimitequ0}
\min_{\hat a\in \mathbb C^k, \hat y_j\in \mathbb R,j=1,\cdots,k}||[\hat \mu ]-[\mu ]||_{\infty} \geq	\min_{\hat a\in \mathbb C^k, \hat y_j\in \mathbb R,j=1,\cdots,k}||\hat B_1 \hat a- B_1 a||_\infty \geq \frac{1}{\sqrt{2n-1}}\min_{\alpha\in \mathbb C^{k}, \hat y_j\in \mathbb R,j=1,\cdots, k}||\hat B_1 \alpha-B_1 a||_2.
	\end{equation}
	
	\textbf{Step 3.}
	 Let $\hat \theta_j = \hat y_j\frac{2r\Omega}{M-1}$. Note that
	\begin{equation}\label{matrixdecomposition1}
	\begin{aligned}
	&\hat B_1=\big(\phi_{2n-2}(e^{i \hat \theta_1}), \cdots,\phi_{2n-2}(e^{i\hat \theta_k} ) \big)\text{diag}(e^{-i\hat y_1\Omega},\cdots,e^{-i\hat y_k\Omega}),\\
	&B_1=\big(\phi_{2n-2}(e^{i \theta_1}), \cdots,\phi_{2n-2}(e^{i \theta_n})\big)\text{diag}(e^{-i y_1\Omega},\cdots,e^{-iy_n\Omega}).
	\end{aligned}
	\end{equation}
We have 
	\begin{align} \label{eq-1111}
	\min_{\alpha\in \mathbb C^{k}, \hat y_j\in \mathbb R,j=1,\cdots, k}||\hat B_1 \alpha-B_1 a||_2= \min_{\alpha\in \mathbb C^{k}, \hat y_j \in \mathbb R,j=1,\cdots, k}||\hat D \alpha-D \tilde{a}||_2,
	\end{align}
	where $\tilde{a}=(a_1e^{-iy_1\Omega},\cdots, a_{n}e^{-iy_n\Omega})^T$, $\hat D=\big(\phi_{2n-2}(e^{i \hat \theta_1}), \cdots,\phi_{2n-2}(e^{i \hat \theta_k} ) \big)$ and $D=\big(\phi_{2n-2}(e^{i \theta_1}), \cdots,\phi_{2n-2}(e^{i \theta_n})\big)$.
	%For 
	%\begin{equation*}
	%\min_{\alpha\in \mathbb C^{k}, \hat y_j\in \mathbb R,j=1,\cdots, k}||\hat D \alpha-D \tilde{a}||_2,
	%\end{equation*}
In view of (\ref{equ:distributionequ1}), we can apply Theorem \ref{spaceapprolowerbound1} to get 
\begin{equation*}
	\min_{\alpha\in \mathbb C^{k}, \hat y_j\in \mathbb R,j=1,\cdots,k}||\hat D \alpha-D \tilde{a}||_2\geq   \frac{m_{\min}\zeta(n)\xi(n-1)(\theta_{\min})^{2n-2}}{\pi^{2n-2}}, 
\end{equation*}
where $\theta_{\min}=\min_{j\neq p}|\theta_j-\theta_p|$. Combing the above estimate with (\ref{equ:upperboundnumberlimitequ0}) and (\ref{eq-1111}), we get
\begin{align}\label{upperboundnumberlimitequ1}
	&\min_{\hat a\in \mathbb C^k, \hat y_j\in \mathbb R,j=1,\cdots,k}||[\hat \mu ]-[\mu ]||_{\infty}\geq 
	\frac{m_{\min}\zeta(n)\xi(n-1)(\theta_{\min})^{2n-2}}{\sqrt{2n-1}\pi^{2n-2}}. 	
\end{align}
\textbf{Step 4.} Recall that $d_{\min}= \min_{j\neq p}|y_j-y_p|$. Using the relation $\theta_j = 	y_j\frac{2r\Omega}{M-1}$ and (\ref{Mdecomposition1}), we can show that 
  \begin{equation*}\label{equ:angleinequ2}
	\theta_{\min}=\frac{2r\Omega}{M-1}d_{\min}\geq \frac{2r\Omega}{(2n-1)(r+1)}d_{\min}\geq \frac{\Omega}{2n-1}d_{\min}. \quad \Big(\text{using $r\geq 1$}\Big) 
\end{equation*}
Then the separation condition (\ref{upperboundnumberlimithm0equ0}) implies
   \begin{equation*}
   \theta_{\min}\geq \frac{4.4 \pi e}{2n-1}\Big(\frac{\sigma}{m_{\min}}\Big)^{\frac{1}{2n-2}}\geq \Big(\frac{2\sqrt{2n-1}}{\zeta(n)\xi(n-1)}\frac{\sigma}{m_{\min}}\Big)^{\frac{1}{2n-2}},
   \end{equation*}
 where we used Lemma \ref{upperboundnumbercalculate1} for the last inequality above. Therefore (\ref{upperboundnumberlimitequ1}) implies that
	\begin{align*}
	&\min_{\hat a\in \mathbb C^k, \hat y_j\in \mathbb R,j=1,\cdots,k}||[\hat \mu ]-[\mu ]||_{\infty} \geq 2\sigma.
	\end{align*}
It follows that
	\begin{align*}
	&||[\hat \mu]-\mathbf Y||_{\infty}=|| [\hat \mu]- [\mu]-\mathbf W||_{\infty}\\
	\geq&||[\hat \mu]- [\mu] ||_{\infty}- ||\mathbf W||_{\infty}\geq|| [\hat \mu ]- [\mu] ||_{\infty}-\sigma \geq \sigma,
	\end{align*}
	which shows that $\hat \mu$ cannot be a $\sigma$-admissible measure. This completes the proof.

	\subsection{Proof of Proposition \ref{numberlowerboundthm0}}
\textbf{Step 1.} Let 
\begin{equation}\label{equ:numberlowerboundequ1}
\tau = \frac{0.81e^{-\frac{3}{2}}}{\Omega}\Big(\frac{\sigma}{m_{\min}}\Big)^{\frac{1}{2n-2}}
\end{equation} 
and $t_1=-(n-1)\tau, t_2=-(n-2)\tau,\cdots, t_n=0, t_{n+1}=\tau,\cdots, t_{2n-1}=(n-1)\tau$. Consider the following system of linear equations  
	\begin{equation}\label{equ:numberlowerboundthm0equ1}
	Aa=0, 
	\end{equation}
	where $A=\big(\phi_{2n-3}(t_1),\cdots,\phi_{2n-3}(t_{2n-1})\big)$. Since $A$ is underdetermined, there exists a nontrivial solution $a=(a_1,\cdots,a_{2n-1})^T$. By the linear independence of the column vectors of $A$, we can show that all $a_j$'s are nonzero. By a scaling of $a$, we can assume that $\min_{1\leq j\leq 2n-1}|a_{j}|=m_{\min}$. We define
\[
\begin{cases}
	\mu=\sum_{j=1}^{n}a_j \delta_{t_j},\  \hat \mu=\sum_{j=n+1}^{2n-1}-a_j\delta_{t_j}, \quad \mbox{if}\,\,\,  \min_{1\leq j\leq n}|a_{j}|=m_{\min},\\
	\mu=\sum_{j=n}^{2n-1}a_j \delta_{t_j},\  \hat \mu=\sum_{j=1}^{n-1}-a_j\delta_{t_j}, \quad  \mbox{otherwise}.
\end{cases}
\]
We shall show that $||[\hat \mu]-[\mu]||_{\infty}<\sigma$ in the subsequent steps. 

\textbf{Step 2.}
Observe that
	\begin{equation}\label{equ:numberlowerboundthm0equ3}
	||[\hat \mu]-[\mu]||_{\infty}\leq \max_{x\in [-\Omega, \Omega]} |\mathcal F(\gamma)(x)|
	\end{equation}
where $\gamma=\sum_{j=1}^{2n-1}a_j \delta_{t_j}$ and 
\begin{equation}\label{Taylorseries1}
	\mathcal F (\gamma)(x)=\sum_{j=1}^{2n-1} a_j e^{it_j x}=\sum_{j=1}^{2n-1}a_j \sum_{k=0}^{\infty}\frac{(it_jx)^k}{k!}=\sum_{k=0}^{\infty}Q_{k}(\gamma)\frac{(ix)^k}{k!}. 
	\end{equation} 
Here $Q_{k}(\gamma)=\sum_{j=1}^{2n-1}a_j t_{j}^k$.
By (\ref{equ:numberlowerboundthm0equ1}), we have 
$Q_{k}(\gamma)=0, k=0,\cdots,2n-3$. We next estimate $Q_{k}(\gamma)$ for $k> 2n-3$.

\textbf{Step 3.}	
We estimate $\sum_{j=1}^{2n-1}|a_{j}|$ first.
We begin by ordering $a_j$'s such that
	\[
	m_{\min}=|a_{j_1}|\leq |a_{j_2}|\leq  \cdots \leq |a_{j_{2n-1}}|.
	\]
Then (\ref{equ:numberlowerboundthm0equ1}) implies that
\[
	a_{j_1}\phi_{2n-3}(t_{j_1}) = \big(\phi_{2n-3}(t_{j_2}), \cdots, \phi_{2n-3}(t_{j_{2n-1}})\big)(-a_{j_2},\cdots, -a_{j_{2n-1}})^T,
\]
and hence
	\[
	a_{j_1} \left(\phi_{2n-3}(t_{j_2}), \cdots, \phi_{2n-3}(t_{j_{2n-1}})\right)^{-1}\phi_{2n-3}(t_{j_1}) = (-a_{j_2},\cdots, -a_{j_{2n-1}})^T.
	\]
	By Lemma \ref{lem:invervandermonde}, we have 
	\[
	a_{j_1}  \Pi_{2\leq q\leq 2n-2}\frac{t_{j_1}-t_{j_q}}{t_{2n-1}-t_{j_q}}= -a_{j_{2n-1}}.
	\]
Therefore
	\[
	|a_{j_{2n-1}}|\leq \frac{(2n-2)!}{(n-1)!(n-2)!}|a_{j_1}|\leq \frac{2^{2n-2}\big((n-1)!\big)^2}{(n-1)!(n-2)!}|a_{j_1}|=2^{2n-2}(n-1)m_{\min},
	\]
and consequently
	\begin{equation}\label{equ:numberlowerboundthm0equ2}
	\sum_{j=1}^{2n-1}|a_j| = 	\sum_{q=1}^{2n-1}|a_{j_q}| \leq (2n-1) |a_{j_{2n-1}}|\leq (2n-1)(n-1)2^{2n-2}m_{\min}.
	\end{equation}
%We next estimate $\max_{x\in [-\Omega, \Omega]} |\mathcal F(\gamma)(x)|$ and it essentially depends on the Taylor series expansion, 
%	\begin{equation}\label{Taylorseries1}
%	\mathcal F (\gamma)(x)=\sum_{j=1}^{2n-1} a_j e^{it_j x}=\sum_{j=1}^{2n-1}a_j \sum_{k=0}^{\infty}\frac{(it_jx)^k}{k!}=\sum_{k=0}^{\infty}Q_{k}(\gamma)\frac{(ix)^k}{k!},
%	\end{equation} 
%	where $Q_{k}(\gamma)=\sum_{j=1}^{2n-1}a_j t_{j}^k$ (see also \cite{akinshin2015accuracy}). 
It follows that for $k\geq 2n-2$,
	\begin{align*}
|Q_{k}(\gamma)|=	|\sum_{j=1}^{2n-1}a_j t_j^{k}|\leq \sum_{j=1}^{2n-1}|a_j| \big((n-1)\tau\big)^{k}\leq (2n-1)(n-1)2^{2n-2}m_{\min} \big((n-1)\tau\big)^{k}.
	\end{align*}
	
\textbf{Step 4.} Using (\ref{Taylorseries1}), we have
	\begin{align*}
	|\max_{x\in [-\Omega, \Omega]}\mathcal F(\gamma)(x)|
	&\leq \sum_{k\geq 2n-2}(2n-1)(n-1)2^{2n-2}m_{\min} \big((n-1)\tau\big)^{k}\frac{\Omega^k}{k!}\\
	&< \frac{2^{2n-2}(2n-1)m_{\min}(n-1)^{2n-1}(\tau\Omega)^{2n-2}}{(2n-2)!}\ \sum_{k=0}^{+\infty}\frac{((n-1)\tau\Omega)^{k}}{k!}\\
	&=\frac{2^{2n-2}(2n-1)m_{\min}(n-1)^{2n-1}(\tau\Omega)^{2n-2}}{(2n-2)!} e^{(n-1)\tau\Omega}\\
	&\leq \frac{(2n-1)\sqrt{n-1}m_{\min}}{2\sqrt{\pi}}(e\tau\Omega)^{2n-2} e^{(n-1)\tau\Omega} \quad \Big(\text{by Lemma \ref{numberlowerboundcalculate1}}\Big)\\
	&\leq \frac{(2n-1)\sqrt{n-1}m_{\min}}{2\sqrt{\pi}}(e\tau\Omega)^{2n-2} e^{(n-1)}. \quad \Big(\text{ noting that (\ref{equ:numberlowerboundequ1}) implies $\tau\Omega\leq 1$}\Big).
\end{align*}	
Finally, using (\ref{equ:numberlowerboundequ1}) and the inequality that $0.81^{2n-2}\frac{(2n-1)\sqrt{n-1}}{2\sqrt{\pi}}<1$, we have
$$
|\max_{x\in [-\Omega, \Omega]}\mathcal F(\gamma)(x)|<\sigma.
$$	
This completes the proof of the proposition.

	\subsection{Proof of Theorem \ref{upperboundsupportlimithm0}}	
	\textbf{Step 1.} Similar to Step 1 in the proof of Theorem \ref{upperboundnumberlimithm0}, we first write
	\begin{equation}\label{Mdecomposition2}
	M=2nr+q, 
	\end{equation}
	where $r, q$ are integers with $r\geq 1$ and $0\leq q< 2n$. It is clear that
		\begin{equation}\label{equ:distributionequ2}
	\theta_j :=y_j\frac{2r\Omega}{M-1}\in \Big[\frac{-\pi}{2}, \frac{\pi}{2}\Big], \quad 	\hat \theta_p :=\hat y_p\frac{2r\Omega}{M-1}\in \Big[\frac{-\pi}{2}, \frac{\pi}{2}\Big], \quad 1\leq j,p\leq n.
	\end{equation}
	Also, by (\ref{Mdecomposition2}),
	\begin{equation}\label{equ:angleinequ4}
	|\theta_j-\theta_p|=\frac{2r\Omega}{M-1}|y_j-y_p|\geq \frac{2r\Omega}{2n(r+1)}\geq  \frac{\Omega}{2n}|y_j-y_p|, \quad \Big(\text{since $r\geq 1$}\Big)
	\end{equation}
	and 
	\begin{equation}\label{equ:angleinequ3}
	\theta_{\min}:=\min_{j\neq p}|\theta_j-\theta_p|\geq \frac{\Omega}{2n}d_{\min}.
	\end{equation}
 \textbf{Step 2.} 
Similar to Step 2 in the proof of Theorem 2.1, 
we consider 
\begin{equation*}
	\left(\mathcal F \hat \mu(\omega_1), \mathcal F \hat \mu(\omega_{1+r}), \cdots,\mathcal F \hat \mu(\omega_{1+(2n-1)r})\right)^T-\left(\mathcal F \mu(\omega_1), \mathcal F \mu(\omega_{1+r}), \cdots,\mathcal F \mu(\omega_{1+(2n-1)r})\right)^T =\hat B_1 \hat a- B_1 a, 
\end{equation*}
where $\hat a= (\hat a_1,\cdots, \hat a_n)^T$, $a=(a_1, \cdots, a_n)^T$ and 
	\begin{equation*}
	\hat B_1= \left(
	\begin{array}{ccc}
	e^{i\hat y_1\omega_1}&\cdots& e^{i\hat y_n\omega_1}\\
	e^{i\hat y_1\omega_{1+r}} &\cdots& e^{i\hat y_n\omega_{1+r}}\\
	\vdots&\vdots&\vdots\\
	e^{i\hat y_1\omega_{1+(2n-1)r}}&\cdots& e^{i\hat y_n \omega_{1+(2n-1)r}}\\
	\end{array}
	\right), \quad B_1=\left(
	\begin{array}{ccc}
	e^{i y_1\omega_1}&\cdots& e^{i y_n\omega_1}\\
	e^{i y_1\omega_{1+r}} &\cdots& e^{i y_n\omega_{1+r}}\\
	\vdots&\vdots&\vdots\\
	e^{i y_1\omega_{1+(2n-1)r}}&\cdots& e^{i y_n \omega_{1+(2n-1)r}}\\
	\end{array}
	\right).
	\end{equation*}
It is clear that 
$$||\hat B_1 \hat a- B_1a||_{\infty}\leq  ||[\hat \mu]- [\mu] ||_\infty.$$
 On the other hand, since $\hat \mu=\sum_{j=1}^n \hat a_j \delta_{\hat y_j}$ is a $\sigma$-admissible measure, we have 
	$||[\hat \mu] -\mathbf Y||_{\infty}<\sigma$.
Therefore 
$$
||[\hat \mu] -[\mu]||_\infty< 2\sigma.
$$
It follows that 
	\begin{equation*}\label{upperboundsupportlimithm1equ0}
	||\hat B_1 \hat a- B_1 a||_\infty < 2\sigma,
	\end{equation*}
	whence we get 
	\begin{equation}\label{upperboundsupportlimithm1equ2}
	||\hat B_1 \hat a- B_1 a||_2 \leq \sqrt{2n}||\hat B_1 \hat a- B_1a||_{\infty}<2\sqrt{2n}\sigma.
	\end{equation}
	
	\textbf{Step 3.} Similar to Step 3 in the proof of Theorem 2.1, we have
%	\[
%	\min_{\alpha\in \mathbb C^n, \hat y_j\in I(n,\Omega),j=1,\cdots, n}||\hat B_1 \alpha-B_1 a||_2.
%	\]
%	Using the same decomposition as  in (\ref{matrixdecomposition1}) for $\hat B_1$ and $B_1$, we have 
	\begin{align}\label{upperboundsupportlimithm1equ1}
	\min_{\alpha\in \mathbb C^n, \hat y_j\in I(n,\Omega),j=1,\cdots, n}||\hat B_1 \alpha-B_1 a||_2= \min_{\alpha\in \mathbb C^n, \hat y_j \in I(n,\Omega),j=1,\cdots, n}||\hat D \alpha-D \tilde{a}||_2
	\end{align}
	where $\tilde{a}=(a_1e^{-iy_1\Omega},\cdots, a_{n}e^{-iy_n\Omega})^T$, $\hat D=\big(\phi_{2n-1}(e^{i \hat \theta_1}),\cdots,\phi_{2n-1}(e^{i \hat \theta_n})\big)$ and $D=\big(\phi_{2n-1}(e^{i \theta_1}),\cdots,\phi_{2n-1}(e^{i \theta_n})\big)$. 
This together with (\ref{upperboundsupportlimithm1equ2}) implies that
$$
\min_{\alpha\in \mathbb C^n, \hat y_j \in I(n,\Omega),j=1,\cdots, n}||\hat D \alpha-D \tilde{a}||_2 \leq 2\sqrt{2n}\sigma. 
$$ 
In view of (\ref{equ:distributionequ2}), we can apply Theorem \ref{spaceapproxlowerbound2} to get
	\begin{align}\label{upperboundsupportlimithm1equ3}
	||\eta_{n,n}(e^{i \theta_1},\cdots,e^{i \theta_n},e^{i \hat \theta_1},\cdots,e^{i \hat \theta_n})||_{\infty}<\frac{\sqrt{2n}2^{n+1}\pi^{n-1}}{\zeta(n)(\theta_{\min})^{n-1}} \frac{\sigma}{m_{\min}}.
	\end{align}
%	where $\theta_{\min}=\min_{p\neq j}|y_p\frac{2r\Omega}{M-1}-y_j\frac{2r\Omega}{M-1}|$. 
%	and 
%	\begin{equation}\label{equ:stableeta1}
%	\eta =\left(
%	\begin{array}{c}
%	|\Pi_{j=1}^n (e^{i\theta_1}_1- e^{i \hat \theta_j})|\\
%	\vdots\\
%	|\Pi_{j=1}^n (e^{i \theta_n}-e^{i \hat \theta_j})|
%	\end{array}
%	\right).
%	\end{equation} 

\textbf{Step 4.}
We apply Corollary \ref{cor:stablemultiproduct1} to estimate $|\hat \theta_j -\theta_j|$'s. For the purpose, let  $\epsilon = \frac{2\sqrt{2n}\pi^{2n-1}}{\zeta(n)(\theta_{\min})^{n-1}} \frac{\sigma}{m_{\min}}$. It is clear that 
$||\eta_{n ,n}||_{\infty}<(\frac{2}{\pi})^n\epsilon$ and we only need to check the following condition
	\begin{equation}\label{upperboundsupportlimithm1equ4}
	\theta_{\min}\geq \Big(\frac{4\epsilon}{\lambda(n)}\Big)^{\frac{1}{n}}, \quad \mbox{or equivalently}\,\,\, (\theta_{\min})^n \geq \frac{4\epsilon}{\lambda(n)}.
	\end{equation}
Indeed, by (\ref{equ:angleinequ3}) and the separation condition (\ref{supportlimithm0equ0}), 
	\begin{equation}\label{upperboundsupportlimithm1equ-1}
	\theta_{\min}\geq  \frac{5.88\pi e}{2n}\Big(\frac{\sigma}{m_{\min}}\Big)^{\frac{1}{2n-1}}\geq  \pi \Big(\frac{8\sqrt{2n}}{\lambda(n)\zeta(n)}\frac{\sigma}{m_{\min}}\Big)^{\frac{1}{2n-1}}, \end{equation}
where we used Lemma \ref{upperboundsupportcalculate1} in the last inequality.  
Then
$$
(\theta_{\min})^{2n-1}\geq \frac{\pi^{2n-1}8\sqrt{2n}}{\lambda(n)\zeta(n)}\frac{\sigma}{m_{\min}}, 
$$
%and it follows that
%$$
%(\theta_{\min})^{n}\geq \frac{\pi^{2n-1}8\sqrt{2n}}{\lambda(n)\zeta(n)(\theta_{\min})^{n-1}},
%$$
whence we get (\ref{upperboundsupportlimithm1equ4}). Therefore, we can apply Corollary \ref{cor:stablemultiproduct1} to get that, after reordering $\hat \theta_j$'s,
	\begin{equation} \label{equ:upperboundsupportlimithm1equ7}
	|\hat \theta_{j}-\theta_j|< \frac{\theta_{\min}}{2}, \text{ and } |\hat \theta_{j}-\theta_j|< \frac{\sqrt{2n}2^{n}\pi^{2n-1}}{\zeta(n)(n-2)!(\theta_{\min})^{2n-2}}\frac{\sigma}{m_{\min}} , j=1,\cdots,n.
	\end{equation}
	
\textbf{Step 5.} Finally, we estimate $|\hat y_j - y_j|$.  Since $|\hat \theta_{j}-\theta_j|< \frac{\theta_{\min}}{2}$, it is clear that 
	$|\hat y_j-y_j|< \frac{d_{\min}}{2}.$
Thus $\hat \mu$ is within the $\frac{d_{\min}}{2}$-neighborhood of $\mu$.
On the other hand, by (\ref{equ:angleinequ4}) 
\[
|\hat y_j-y_j|\leq \frac{2n}{\Omega}|\hat \theta_j -\theta_j|. 
\]	
Using (\ref{equ:upperboundsupportlimithm1equ7}), (\ref{equ:angleinequ3}) and Lemma \ref{upperboundsupportcalculate2}, a direct calculation shows that 
% and (\ref{equ:upperboundsupportlimithm1equ7}), we have
%	\[|\hat y_j-y_j|\leq \frac{n+\frac{1}{2}}{\Omega}|\hat \theta_j -\theta_j|<\frac{n+\frac{1}{2}}{\Omega} \frac{\sqrt{2n}2^{n}\pi^{2n-1}}{\zeta(n)(n-2)!(\theta_{\min})^{2n-2}}\frac{\sigma}{m_{\min}}, \quad j=1,\cdots,n.\]
%employing (\ref{equ:angleinequ3}) and Lemma \ref{upperboundsupportcalculate2}, 
%we can calculate that 
	\begin{align*}
	|\hat y_j-y_j|< \frac{C(n)}{\Omega} (\frac{\pi}{\Omega d_{\min}})^{2n-2} \frac{\sigma}{m_{\min}}, 
	\end{align*}
	where $C(n)=n2^{4n-2}e^{2n}\pi^{-\frac{1}{2}}$. This completes the proof.
	
	\subsection{Proof of Proposition \ref{supportlowerboundthm0}}
	Proof: Let $t_1=-n\tau,t_2=-(n-1)\tau,\cdots, t_n=-\tau, t_{n+1}=0,\cdots, t_{2n}=(n-1)\tau$. Consider the following system of linear equations 
	\begin{equation}
	Aa=0,
	\end{equation}
	where $A=\big(\phi_{2n-2}(t_1),\cdots,\phi_{2n-2}(t_{2n})\big)$ with $\phi_{2n-2}(t)$ defined in (\ref{phiformula}). 
%	Since $A$ is underdetermined, we have non-zero $a=(a_1,\cdots,a_{2n})$ satisfying the above equation. By the linear independence of the column vectors in the matrix $A$, we can show that all $a_j$'s are not zero.
Similar to the argument in the proof of Proposition \ref{numberlowerboundthm0}, we can show that after a scaling of $a$,
\begin{equation}\label{equ:supportlowerboundthm0equ1}
	m_{\min} =\min_{1\leq j \leq 2n}|a_j|, \quad	|\sum_{j=1}^{2n}a_j|\leq n^22^{2n}m_{\min}.
\end{equation}
We define $\mu=\sum_{j=1}^{n}a_j \delta_{t_j},\  \hat \mu=\sum_{j=n+1}^{2n}-a_j\delta_{t_j}$. 
We prove that $$||[\hat \mu]-[\mu]||_{\infty}\leq \max_{x\in[-\Omega,\Omega]}|\mathcal F (\gamma)(x)|<\sigma,$$
where $\gamma=\sum_{j=1}^{2n}a_j \delta_{t_j}$ and $\mathcal F (\gamma)$ is defined as in (\ref{Taylorseries1}).
% by estimating $\max_{x\in[-\Omega, \Omega]}|\mathcal F (\gamma)(x)|$ for $\gamma=\sum_{j=1}^{2n}a_j \delta_{t_j}$. 
Indeed, (\ref{equ:supportlowerboundthm0equ1}) implies, for $k\geq 2n-1$,
	\begin{align*}
	|\sum_{j=1}^{2n}a_j t_j^{k}|\leq \sum_{j=1}^{2n}|a_j| (n\tau)^{k}\leq n^22^{2n}m_{\min}  (n\tau)^{k}.
	\end{align*}
On the other hand, using (\ref{Taylorseries1}), we have 
	\[Q_{k}(\gamma)=0, k=0,\cdots,2n-2,\ {\text{and}}\ |Q_{k}(\gamma)|\leq n^22^{2n}m_{\min}(n\tau)^{k}, k\geq 2n-1. \]
	Therefore, for $|x|\leq \Omega$,
	\begin{align*}
	&|\mathcal F (\gamma)(x)|\leq \sum_{k\geq 2n-1}n^22^{2n}m_{\min} (n\tau)^{k}\frac{|x|^k}{k!}\leq \sum_{k\geq 2n-1}n^22^{2n}m_{\min} (   n \tau)^{k}\frac{\Omega^k}{k!}\\
	<& \frac{2^{2n}m_{\min}n^{2n+1}(\tau\Omega)^{2n-1}}{(2n-1)!}\ \sum_{k=0}^{+\infty}\frac{(n\tau\Omega)^{k}}{k!}=\frac{2^{2n}m_{\min}n^{2n+1}(\tau\Omega)^{2n-1}}{(2n-1)!} e^{n\tau\Omega}\\
	\leq& \frac{2^{2n}m_{\min}n^{2n+1}(\tau\Omega)^{2n-1}}{(2n-1)!} e^{n} \quad \Big(\text{(\ref{supportlowerboundequ0}) implies $\tau \Omega \leq 1 $}\Big)\\
	\leq& \frac{m_{\min}e^{\frac{3}{2}}n^2}{\sqrt{\pi(n-\frac{1}{2})}}e^{3n-\frac{3}{2}}(\tau\Omega)^{2n-1}\quad \Big(\text{by Lemma \ref{numberlowerboundcalculate1}}\Big)\\
	<& \sigma.  \quad \Big(\text{by (\ref{supportlowerboundequ0}) and $0.49^{2n-1} \frac{e^{\frac{3}{2}}n^2}{\sqrt{\pi(n-\frac{1}{2})}}<1$} \Big)
	\end{align*} 
It follows that $||[\hat \mu]-[\mu]||_{\infty}<\sigma$.

	\section{A sweeping singular-value-thresholding number detection algorithm and phase transition}\label{section:numberalgorithm}
	
We propose a number detection algorithm in this section and verify the phase transition phenomenon in the number detection of the LSE problem. In the statistical setting with multiple measurements (multiple $\mathbf Y$'s), the number of spectra or the model order is usually determined by two approaches. One approach selects the model which includes the model order using some generic information theoretic criteria by minimizing the summation of a log-likelihood function and a regularization term of the free parameters in the model. Examples include AIC \cite{akaike1998information, akaike1974new, wax1985detection}, BIC/MDL \cite{schwarz1978estimating, rissanen1978modeling, wax1989detection}. The other approach determines the model order by thresholding the eigenvalues of the covariance matrix of the data (the so-called eigen-thresholding method), see for instance \cite{lawley1956tests, chen1991detection, he2010detecting, han2013improved}.

In this paper, we deal with the case of deterministic noise with a single measurement, see (\ref{equ:modelsetting1}).  Following the idea of eigen-thresholding, 
we derive a deterministic threshold to the singular values of the Hankel matrix formed from the measurement data (see Theorem \ref{MUSICthm1}),  and use this threshold to estimate the number of line spectra. We term this algorithm singular-value-thresholding algorithm.

To be more specific, we choose partial measurement at the sample points $z_t= \omega_{(t-1)r+1}$ for $t=1,\cdots,2s+1$ where $s\geq n$ and $r=(M-1) \mod 2s$. 
%instead of considering the full measurement $\mathbf Y = (\mathbf Y(\omega_1),\cdots, \mathbf Y(\omega_M))^T$,  
For ease of exposition, we assume $r=\frac{M-1}{2s}$. Then $z_t=\omega_{(t-1)\frac{M-1}{2s}+1} =-\Omega+\frac{t-1}{s}\Omega$ (since $\omega_1=-\Omega$, $\omega_{M}=\Omega$) and the partial measurement is
	\[
	\mathbf Y(z_t)= \mathcal F \mu (z_t) + \mathbf W(z_t)= \sum_{j=1}^{n}a_j e^{i y_j z_t} +\mathbf W(z_t), \quad 1\leq t \leq 2s+1.
	\]
We form the following Hankel matrix
	\begin{equation}\label{hankelmatrix1}
	\mathbf H(s)=\left(\begin{array}{cccc}
	\mathbf Y(-\Omega)&\mathbf Y(-\Omega+\frac{1}{s}\Omega)&\cdots& \mathbf Y(0)\\
	\mathbf Y(-\Omega+\frac{1}{s}\Omega)&\mathbf Y(-\Omega+\frac{2}{s}\Omega)&\cdots&\mathbf Y(\frac{1}{s}\Omega)\\
	\cdots&\cdots&\ddots&\cdots\\
	\mathbf Y(0)&\mathbf Y(\frac{1}{s}\Omega)&\cdots&\mathbf Y(\Omega)
	\end{array}
	\right).\end{equation}
	We observe that $\mathbf H(s)$ has the decomposition
	\[\mathbf H(s)= DAD^T+\Delta,\]
	where $A=\text{diag}(e^{-iy_1\Omega}a_1, \cdots, e^{-iy_n\Omega}a_n)$ and $D=\big(\phi_{s}(e^{i y_1 \frac{\Omega}{s}}), \cdots, \phi_{s}(e^{i y_n \frac{\Omega}{s}})\big)$ with $\phi_{s}(\omega)$ being defined in (\ref{phiformula}) and 
	\begin{equation*}
	\Delta = \left(\begin{array}{cccc}
	\mathbf {W}(-\Omega)&\mathbf {W}(-\Omega+\frac{1}{s}\Omega)&\cdots& \mathbf {W}(0)\\
	\mathbf {W}(-\Omega+\frac{1}{s}\Omega)&\mathbf {W}(-\Omega+\frac{2}{s}\Omega)&\cdots&\mathbf {W}(\frac{1}{s}\Omega)\\
	\vdots&\vdots&\ddots&\vdots\\
	\mathbf {W}(0)&\mathbf {W}(\frac{1}{s}\Omega)&\cdots&\mathbf {W}(\Omega)
	\end{array}
	\right).
	\end{equation*}
	We denote the singular value decomposition of $\mathbf H(s)$ as  
	\[\mathbf H(s)=\hat U\hat \Sigma \hat U^*,\]
	where $\hat\Sigma =\text{diag}(\hat \sigma_1,\cdots, \hat \sigma_n, \hat \sigma_{n+1},\cdots,\hat\sigma_{s+1})$ with the singular values $\hat \sigma_j$, $1\leq j \leq s+1$, ordered in a decreasing manner. Note that when there is no noise, $\mathbf H(s)=DAD^T$. We have the following estimate for the singular values of $DAD^T$.
	\begin{lem}\label{MUSICthm0}
		Let $n\geq 2$, $y_j\in I(n,\Omega), 1\leq j\leq n,$ and let $\sigma_1,\cdots,\sigma_n,0,\cdots,0$ be the singular values of $DAD^T$ ordered in a decreasing manner. Then the following estimate holds
		\begin{equation}\label{MUSICthm0equ1}
		\sigma_n\geq \frac{m_{\min}\zeta(n)^2\big(\theta_{\min}(\Omega,s)\big)^{2n-2}}{n\pi^{2n-2}},
		\end{equation}
		where $\zeta(n)$ is defined in (\ref{zetaxiformula1}) and $\theta_{\min}(\Omega,s)=\min_{p\neq j}|y_p\frac{\Omega}{s}-y_j \frac{\Omega}{s}|$.
	\end{lem}
	Proof: Recall that $\sigma_n$ is the minimum nonzero singular value of $DAD^T$. Let $\ker(D^T)$ be the kernel space of $D^T$ and $\ker^{\perp}(D^T)$ be its orthogonal complement, we have 
	\begin{align*}
	\sigma_n=\min_{||x||_2=1,x\in \ker^{\perp}(D^T)}||DAD^Tx||_2\geq \sigma_{\min}(DA)\sigma_n(D^T)\geq \sigma_{\min}(D)\sigma_{\min}(A)\sigma_{\min}(D).
	\end{align*}
	Since $s\geq n$, $y_j\frac{\Omega}{s}\in \big[\frac{-\pi}{2}, \frac{\pi}{2}\big]$ for $y_j\in I(n,\Omega)=\big[-\frac{(n-1)\pi}{2\Omega}, \frac{(n-1)\pi}{2\Omega}\big]$. By Lemma \ref{singularvaluevandermonde2} and  \ref{norminversevandermonde2}, we have
	\begin{align*}
	\sigma_{\min}(D)\geq \frac{1}{\sqrt{n}}\frac{\zeta(n)\big(\theta_{\min}(\Omega,s)\big)^{n-1}}{\pi^{n-1}}.
	\end{align*}
	It follows that
	\[\sigma_n\geq \sigma_{\min}(A)\Big(\frac{1}{\sqrt{n}}\frac{\zeta(n)\big(\theta_{\min}(\Omega,s)\big)^{n-1}}{\pi^{n-1}}\Big)^2\geq \frac{m_{\min}\zeta(n)^2\big(\theta_{\min}(\Omega,s)\big)^{2n-2}}{n\pi^{2n-2}}.\]
	
	\vspace{0.3cm}
We next present the main result on the threshold for the singular values of the matrix $\mathbf{H}(s)$.
	\begin{thm}\label{MUSICthm1}
		Let $\sigma<m_{\min}, s\geq n$ and $\mu=\sum_{j=1}^{n}a_j \delta_{y_j}$ with $y_j\in I(n,\Omega), 1\leq j\leq n$. We have 
		\begin{equation}\label{MUSICthm1equ-1}
		\hat \sigma_j\leq  (s+1)\sigma,\quad j=n+1,\cdots,s+1.
		\end{equation}
		Moreover, if the following separation condition is satisfied
		\begin{equation}\label{MUSICthm1equ0}
		\min_{p\neq j}|y_p-y_j|>\frac{\pi s}{\Omega}\Big(\frac{2n(s+1)}{\zeta(n)^2}\frac{\sigma}{m_{\min}}\Big)^{\frac{1}{2n-2}},
		\end{equation}
		then
		\begin{equation}\label{MUSICthm1equ2}
		\hat\sigma_{n}>(s+1)\sigma.
		\end{equation}
	\end{thm}
	Proof: Since $||\mathbf {W}||_{\infty}\leq \sigma$, we have $||\Delta||_2\leq ||\Delta||_F\leq (s+1)\sigma$. By Weyl's theorem, we have $|\hat \sigma_j-\sigma_j|\leq ||\Delta||_2, j=1,\cdots,n$. Together with $\sigma_j=0, n+1\leq j \leq s+1$, we get $|\hat \sigma_j|\leq ||\Delta||_2\leq(s+1)\sigma, n+1\leq j \leq s+1$. This proves (\ref{MUSICthm1equ-1}). 
	
	Let $\theta_{\min}(\Omega,s)=\min_{p\neq j}\Big|y_p\frac{\Omega}{s}-y_j\frac{\Omega}{s}\Big|=\frac{\Omega}{s}\min_{p\neq j}\Big|y_p-y_j\Big|$. The separation condition (\ref{MUSICthm1equ0}) implies $\theta_{\min}(\Omega,s) > \pi \Big(\frac{2n(s+1)}{\zeta(n)^2}\frac{\sigma}{m_{\min}}\Big)^{\frac{1}{2n-2}}$.
	By (\ref{MUSICthm0equ1}), we have 
	\begin{align}\label{MUSICthm1equ7}
	\sigma_n \geq\frac{m_{\min}\zeta(n)^2\theta_{\min}(\Omega,s)^{2n-2}}{n\pi^{2n-2}}>2(s+1)\sigma.
	\end{align}
Similarly, by Weyl's theorem, $|\hat \sigma_n-\sigma_n|\leq ||\Delta||_2$. Thus, $\hat \sigma_n> 2(s+1)\sigma-||\Delta||_2\geq (s+1)\sigma$. Conclusion (\ref{MUSICthm1equ2}) follows. 
	
\medskip
	
Based on Theorem \ref{MUSICthm1}, we can propose a simple thresholding algorithm, \textbf{Algorithm 1}, for the number detection.
% where we also use a decimation strategy implemented in \cite{liu2019computational, batenkov2018stability}.
	
	\begin{algorithm}
		\caption{\textbf{Singular-value-thresholding number detection algorithm}}
		\textbf{Input:} Number $s$, Noise level $\sigma$\\
		\textbf{Input:} measurement: $\mathbf{Y}=(\mathbf Y(\omega_1),\cdots, \mathbf Y(\omega_M))^T$\\	
		1: $r=(M-1)\mod 2s$,  $\mathbf{Y}_{new}=(\mathbf Y(\omega_1), \mathbf Y(\omega_{r+1}), \cdots, \mathbf Y(\omega_{2sr+1}))^T$\;
		2: Formulate the $(s+1)\times(s+1)$ Hankel matrix $\mathbf H(s)$ from $\mathbf{Y}_{new}$, and
		compute the singular value of $\mathbf H(s)$ as $\hat \sigma_{1}, \cdots,\hat \sigma_{s+1}$ distributed in a decreasing manner\;
		4: Determine $n$ by $\hat \sigma_n>(s+1)\sigma$ and $\hat \sigma_{j}\leq (s+1)\sigma, j=n+1,\cdots, s+1$\;
		\textbf{Return:} $n$ 
	\end{algorithm}
Note that for \textbf{Algorithm 1} to work, in addition to the noise level $\sigma$, we also need the integer $s$ which is required to be greater than the number of line spectra. However, a suitable $s$ is not easy to estimate and large $s$ may incur a deterioration of resolution as indicated by (\ref{MUSICthm1equ0}). To remedy the issue, we propose a sweeping singular-value-thresholding number detection algorithm (\textbf{Algorithm 2}) below.	In short, we detect the number by \textbf{Algorithm 1} for all $s$ from $1$ to $\lfloor \frac{M-1}{2}\rfloor$, and choose the greatest one as the number of spectra. The idea of \textbf{Algorithm2} is as follows. We know that when $s=n$ and the line spectra satisfy 
\begin{equation}\label{equ:numberalgorithmequ1}
\min_{p\neq j}|y_j-y_p| \geq \frac{n\pi }{\Omega}\Big(\frac{2n(n+1)}{\zeta(n)^2}\frac{\sigma}{m_{\min}}\Big)^{\frac{1}{2n-2}},
\end{equation}
Theorem \ref{MUSICthm1} implies \textbf{Algorithm 1} can exactly detect the number $n$ when $s=n$. As $s$ increases to values greater than $n$, (\ref{MUSICthm1equ-1}) implies that the number detected by \textbf{Algorithm 1} will not exceed $n$. Therefore, the sweeping singular-value-thresholding algorithm can detect the exact number $n$ when $M$ is greater than $2n+1$ and the line spectra are separated more than $\frac{n\pi }{\Omega}\Big(\frac{2n(n+1)}{\zeta(n)^2}\frac{\sigma}{m_{\min}}\Big)^{\frac{1}{2n-2}}$.   
We note that by Lemma \ref{musiccalculate1}, $ \frac{n\pi }{\Omega}\Big(\frac{2n(n+1)}{\zeta(n)^2}\Big)^{\frac{1}{2n-2}} \leq \frac{3\pi e}{\Omega}$. So the minimum separation distance required for \textbf{Algorithm 2} to work is comparable to the upper bound we derived in (\ref{upperboundnumberlimithm0equ0}).

%
%\begin{equation}\label{equ:numberalgorithmequ1}
%\min_{p\neq j}|y_j-y_p|\geq \frac{3\pi e}{\Omega}\Big(\frac{\sigma}{m_{\min}}\Big)^{\frac{1}{2n-2}} \geq \frac{n\pi }{\Omega}\Big(\frac{2n(n+1)}{\zeta(n)^2}\frac{\sigma}{m_{\min}}\Big)^{\frac{1}{2n-2}},
%\end{equation}
%where we used Lemma \ref{musiccalculate1} for the last inequality above, 
%%(\ref{MUSICthm1equ0}), 
%(\ref{MUSICthm1equ2}) holds. Therefore, under separation condition (\ref{equ:numberalgorithmequ1}), Theorem \ref{MUSICthm1} implies \textbf{Algorithm 1} can exactly detect the number $n$ when $s=n$. As $s$ increases to values greater than $n$, (\ref{MUSICthm1equ-1}) implies that the number detected by \textbf{Algorithm 1} will not exceed $n$. Therefore, the sweeping singular-value-thresholding algorithm can detect the exact number $n$ when the line spectra are separated more than $\frac{3\pi e}{\Omega}\Big(\frac{\sigma}{m_{\min}}\Big)^{\frac{1}{2n-2}}$ and $M$ is greater than $2n+1$. 

	\medskip
	
	\begin{algorithm}[H]
		\caption{\textbf{Sweeping singular-value-thresholding number detection algorithm}}	
		\textbf{Input:} Noise level $\sigma$, measurement: $\mathbf{Y}=(\mathbf Y(\omega_1),\cdots, \mathbf Y(\omega_M))^T$\\
		\textbf{Input:} $n_{max}=0$\\
		\For{$s=1: \lfloor \frac{M-1}{2}\rfloor$}{
			Input $s,\sigma, \mathbf{Y}$ to \textbf{Algorithm 1}, save the output of \textbf{Algorithm 1} as $n_{recover}$\; 
			\If{$n_{recover}>n_{max}$}{$n_{max}=n_{recover}$}
		}
		{
			\textbf{Return} $n_{max}$
		}
	\end{algorithm}

	\medskip
	We next conduct numerical experiments to demonstrate the efficiency of \textbf{Algorithm 2}. 
	
	\textbf{Experiment 1:} We set $n=4,\Omega=1,\sigma=1\times 10^{-7}$ and 
	\[\mu=\delta_{y_1}-\delta_{y_2}-\delta_{y_3}+\delta_{y_4},\]
	where $y_1=-0.5, y_2=0, y_3=0.5,y_4=1$. We measure at $M=20$ sample points evenly spaced in $[-\Omega, \Omega]$. The noisy measurement is
	\[
	\mathbf Y= \big(\mathbf Y(\omega_1),\mathbf Y(\omega_2),\cdots,\mathbf Y(\omega_M)\big)^T,
	\]
	where $\mathbf Y(\omega_q)=\sum_{j=1}^{n}a_je^{i y_j \omega_q}+\mathbf W(\omega_q)$ with $\mathbf W(\omega_q)$ the noise satisfying $||\mathbf W||_{\infty}<\sigma$. Using \textbf{Algorithm 2} we can recover the number $n=4$.
	
\begin{figure}[!h]
		\centering
		\includegraphics[width=2.5 in]{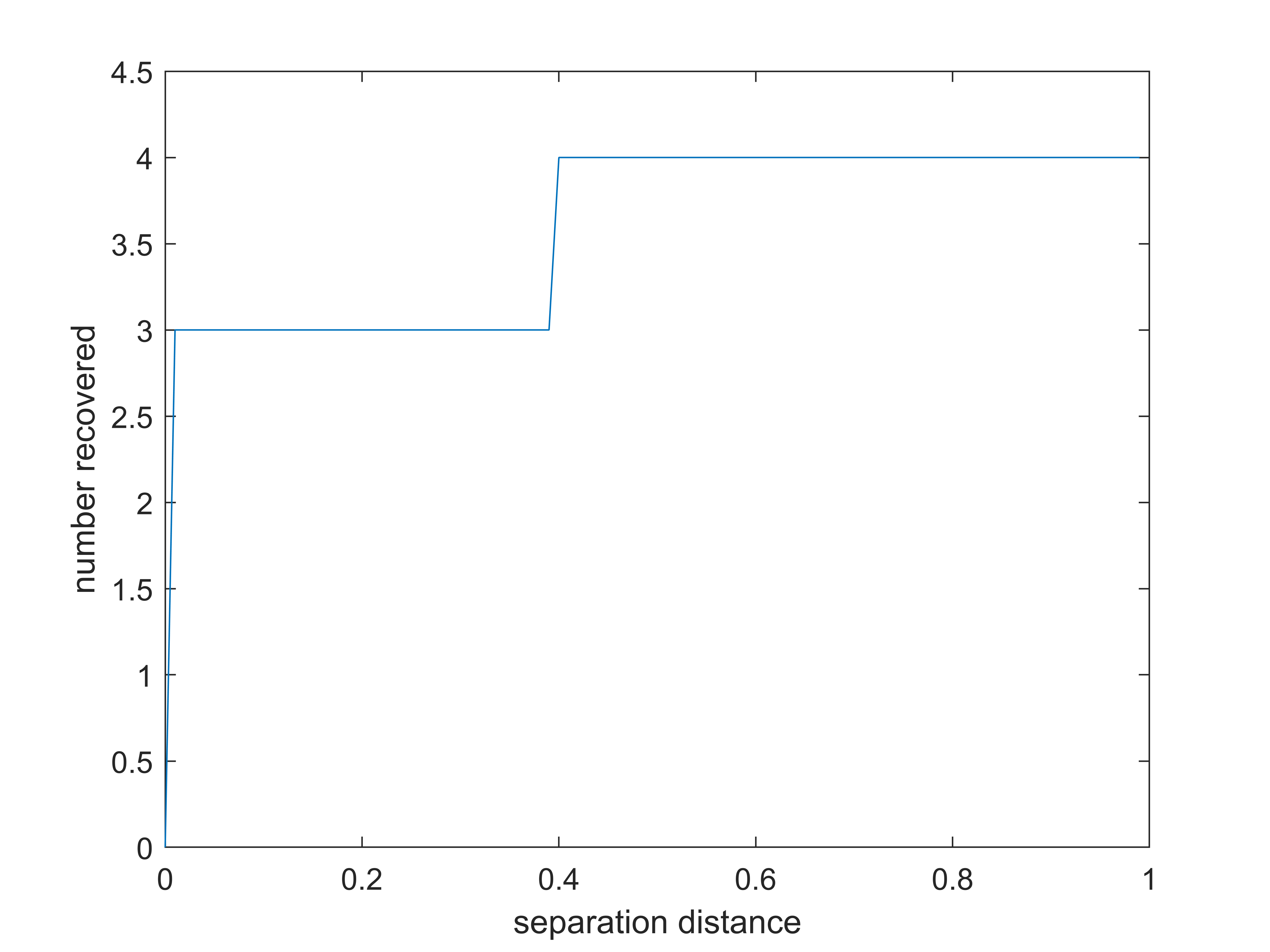}
		\caption{number detected by \textbf{Algorithm 2} w.r.t separation distance}
		\label{fig:MUISCdistancesig1}
	\end{figure}

	We then apply the algorithm to line spectra with different separation distances to find the minimum separation distance required for the success of the algorithm. Precisely, we set $y_1=-\tau, y_2=0, y_3=\tau, y_4=2\tau$, and detect the number by \textbf{Algorithm 2} as $\tau$ varies from $0$ to $1$. We plot Figure \ref{fig:MUISCdistancesig1} which illustrates the number detected when this minimum separation distance varies.
	It shows that we can exactly recover the spectral number when they are separated more than 0.41 ($\approx 0.13\frac{\pi}{\Omega}$).

	\subsection{Phase transition} \label{sec-phase}
	We know from Section \ref{section:Mainresults} that the resolution limit to the number detection problem is bounded from below and above by  $\frac{C_1}{\Omega}(\frac{\sigma}{m_{\min}})^{\frac{1}{2n-2}}$ and $\frac{C_2}{\Omega}(\frac{\sigma}{m_{\min}})^{\frac{1}{2n-2}}$ respectively for some constants $C_1,C_2$. We shall demonstrate that this implies  
	a phase transition phenomenon for the number detection.  
	Precisely, recall that the super-resolution factor is 
	$SRF=\frac{\pi}{d_{\min} \Omega}$ and the signal-to-noise ratio is $SNR= \frac{m_{\min}}{\sigma}$. From the two bounds for the resolution limit, we can draw the conclusion that exact number detection is guaranteed if
	$$
	\log(SNR) > (2n-2)\log(SRF)+(2n-2) \log \frac{C_1}{\pi}, 
	$$
	and may fail if
	$$
	\log(SNR) < (2n-2)\log(SRF)+(2n-2) \log \frac{C_2}{\pi}.  
	$$
	As a consequence, we can see that in the parameter space of $\log SNR-\log SRF$, there exist two lines both with slope $2n-2$ such that the number detection is successful for cases above the first line and unsuccessful for cases below the second. In the intermediate region between the two lines, the number detection can be either successful or unsuccessful from case to case. This is clearly demonstrated in the numerical experiments below. 
	
\begin{figure}[!h]
	\centering
\begin{subfigure}[b]{0.28\textwidth}
	\centering
	\includegraphics[width=\textwidth]{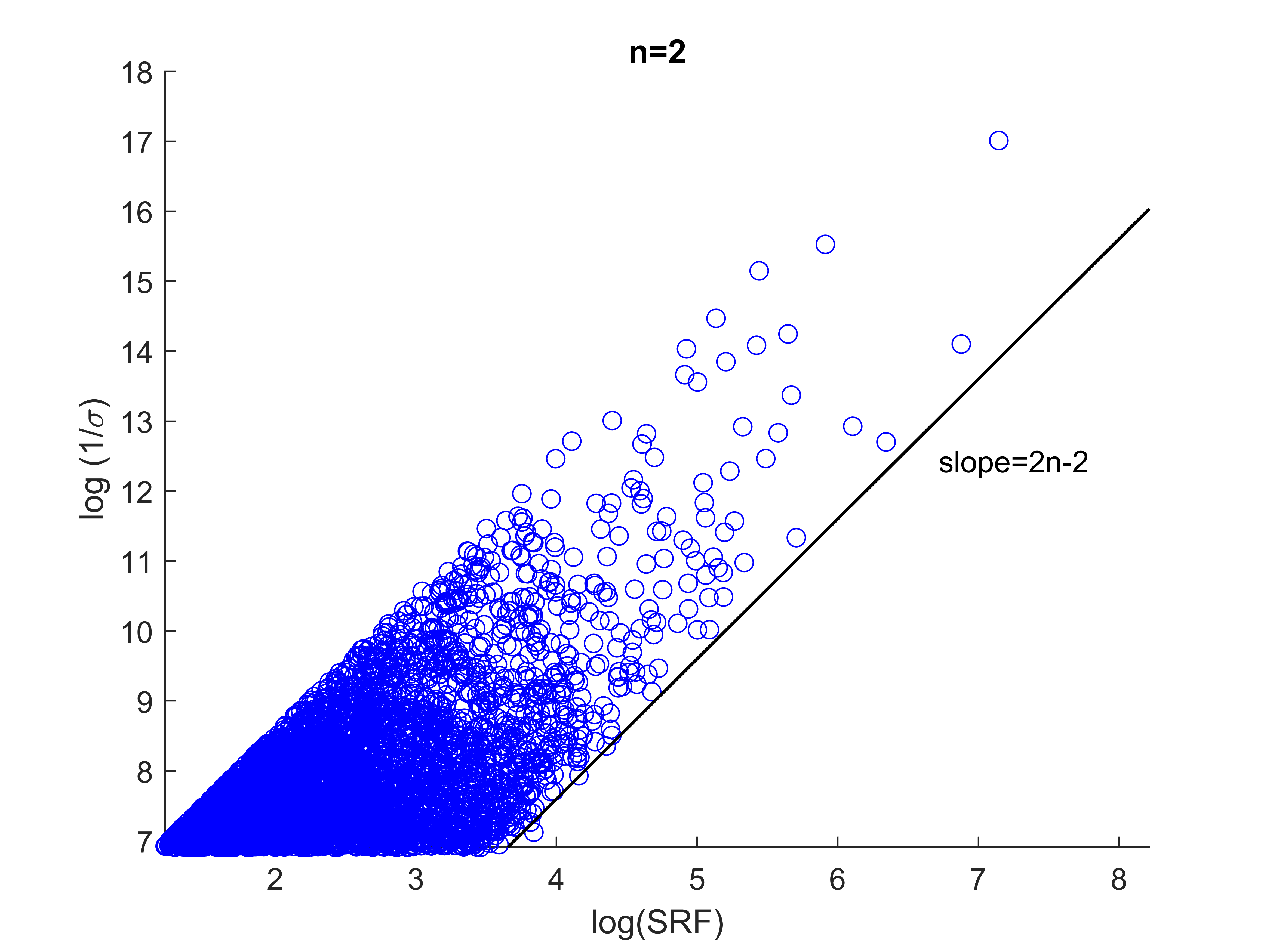}
	\caption{detection success}
\end{subfigure}
\begin{subfigure}[b]{0.28\textwidth}
	\centering
	\includegraphics[width=\textwidth]{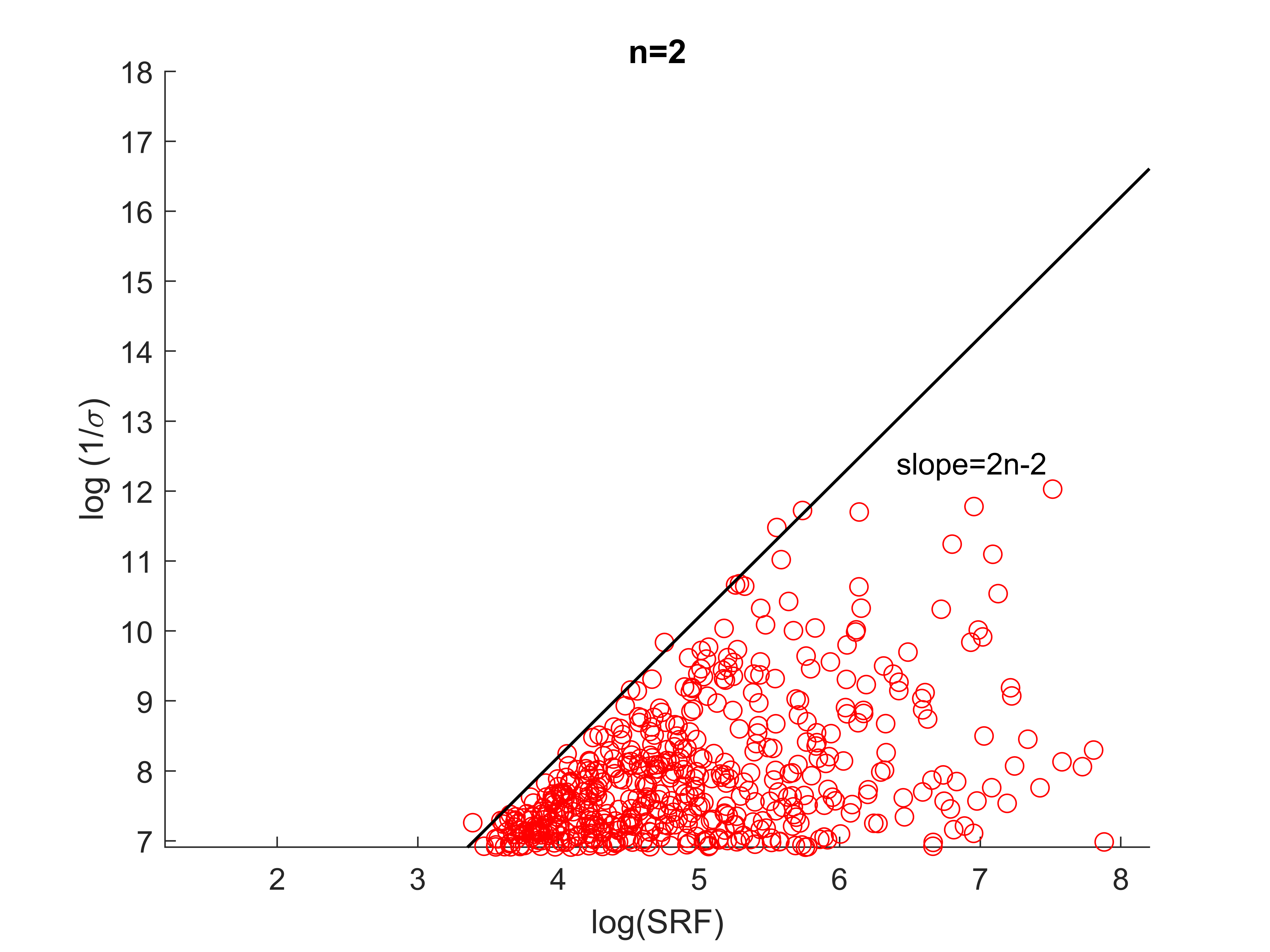}
	\caption{detection fail}
\end{subfigure}
\begin{subfigure}[b]{0.28\textwidth}
	\centering
	\includegraphics[width=\textwidth]{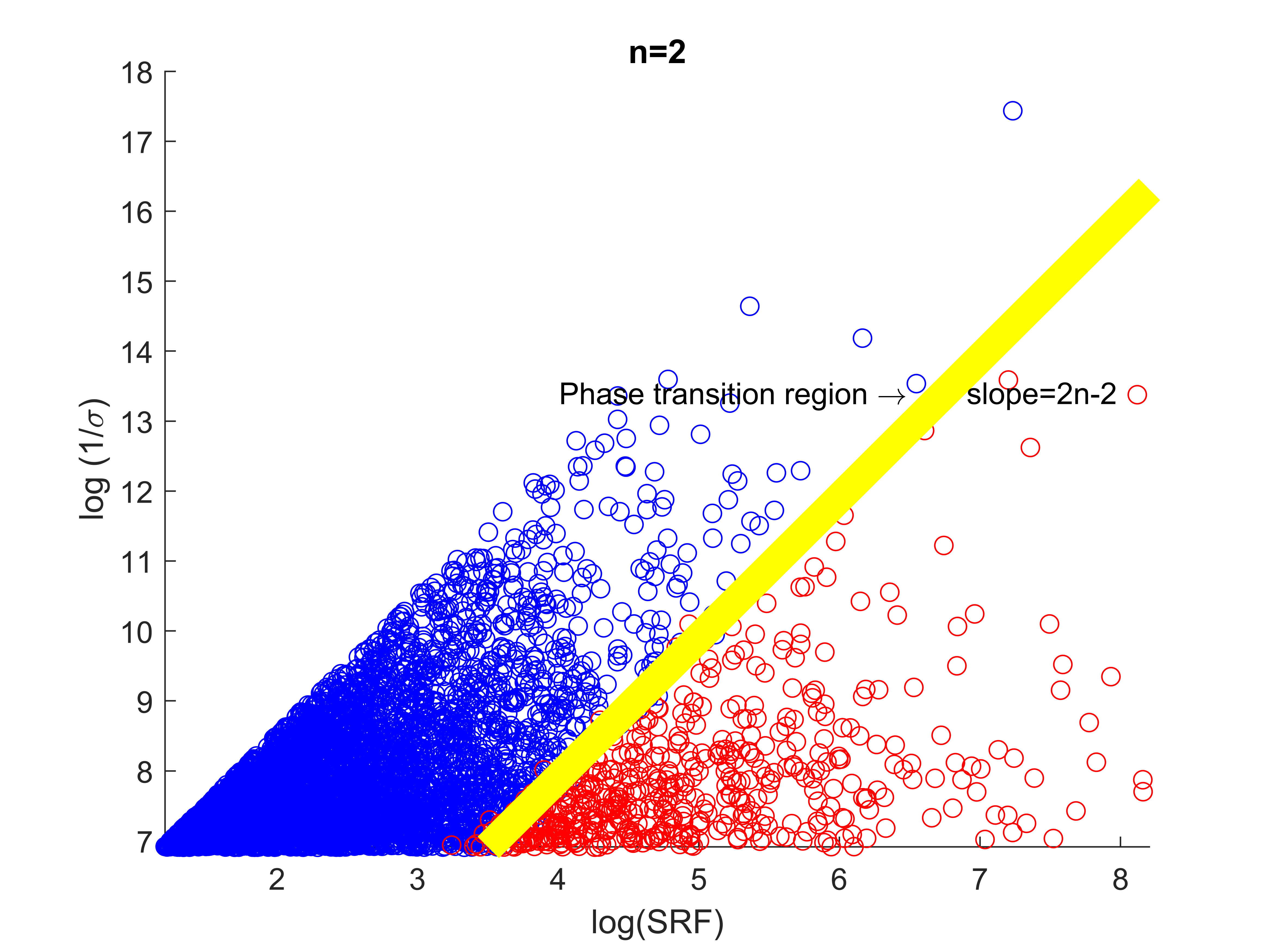}
	\caption{phase transition region}
\end{subfigure}
\begin{subfigure}[b]{0.28\textwidth}
	\centering
	\includegraphics[width=\textwidth]{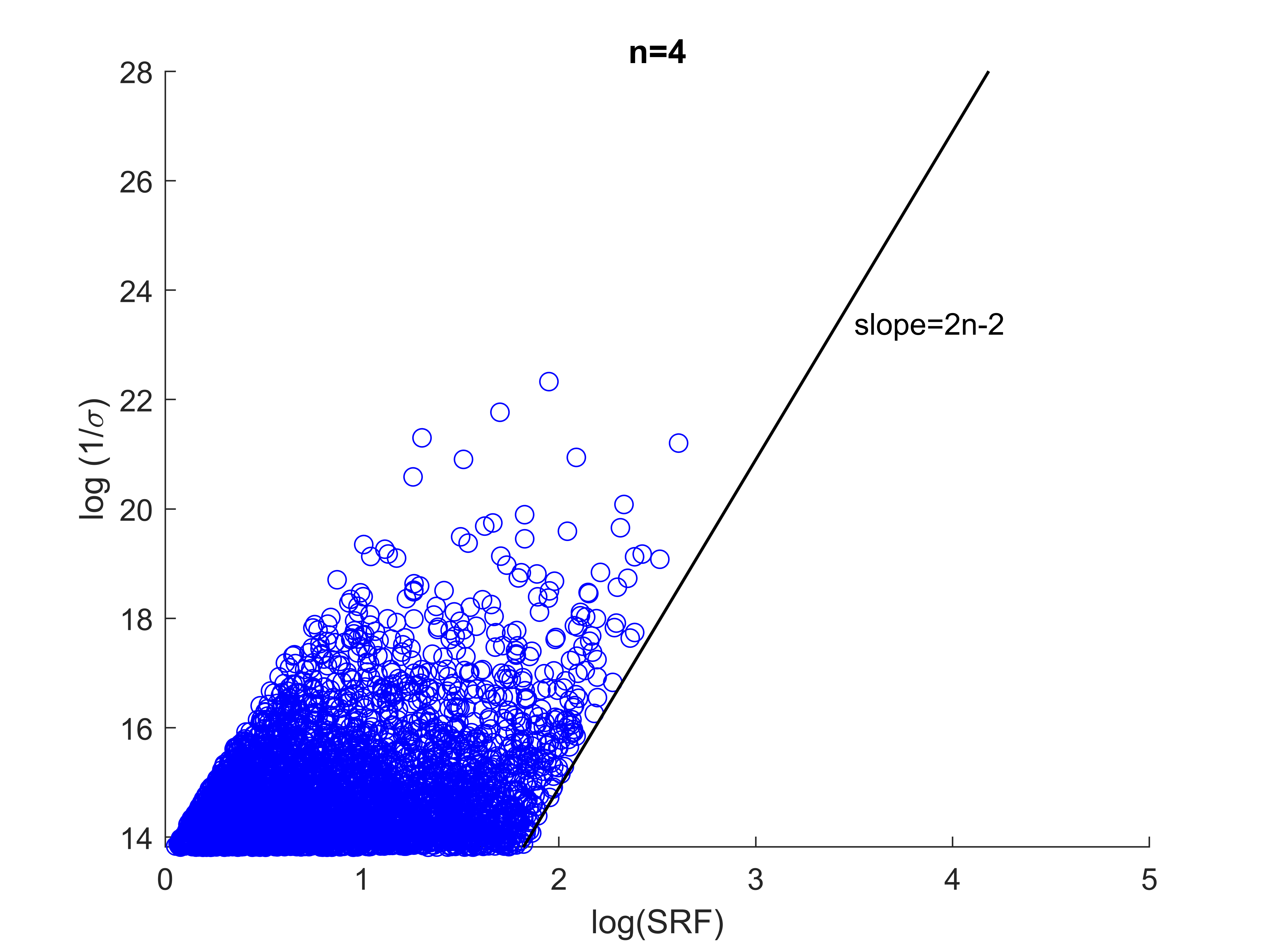}
	\caption{detection success}
\end{subfigure}
\begin{subfigure}[b]{0.28\textwidth}
	\centering
	\includegraphics[width=\textwidth]{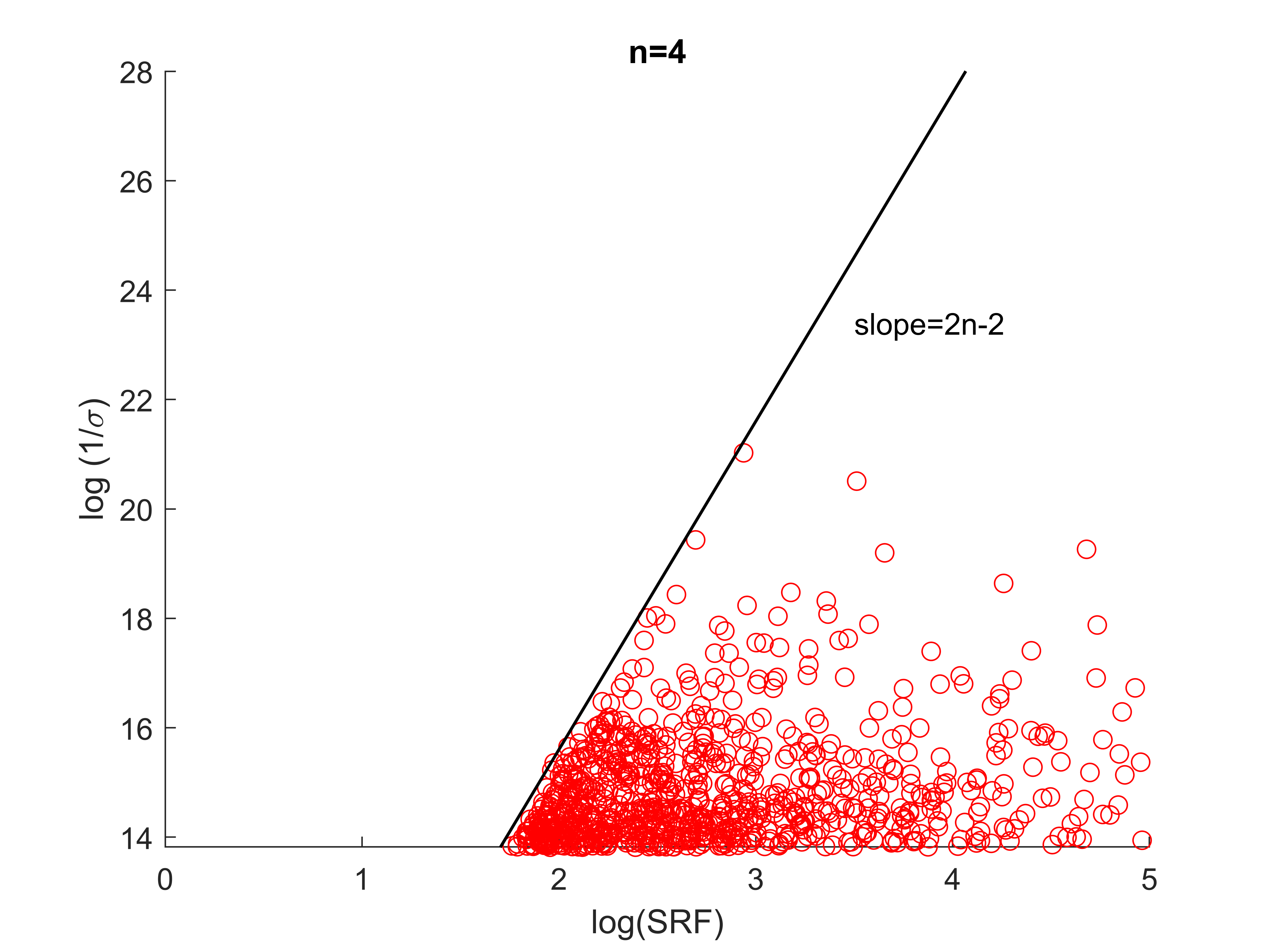}
	\caption{detection fail}
\end{subfigure}
\begin{subfigure}[b]{0.28\textwidth}
	\centering
	\includegraphics[width=\textwidth]{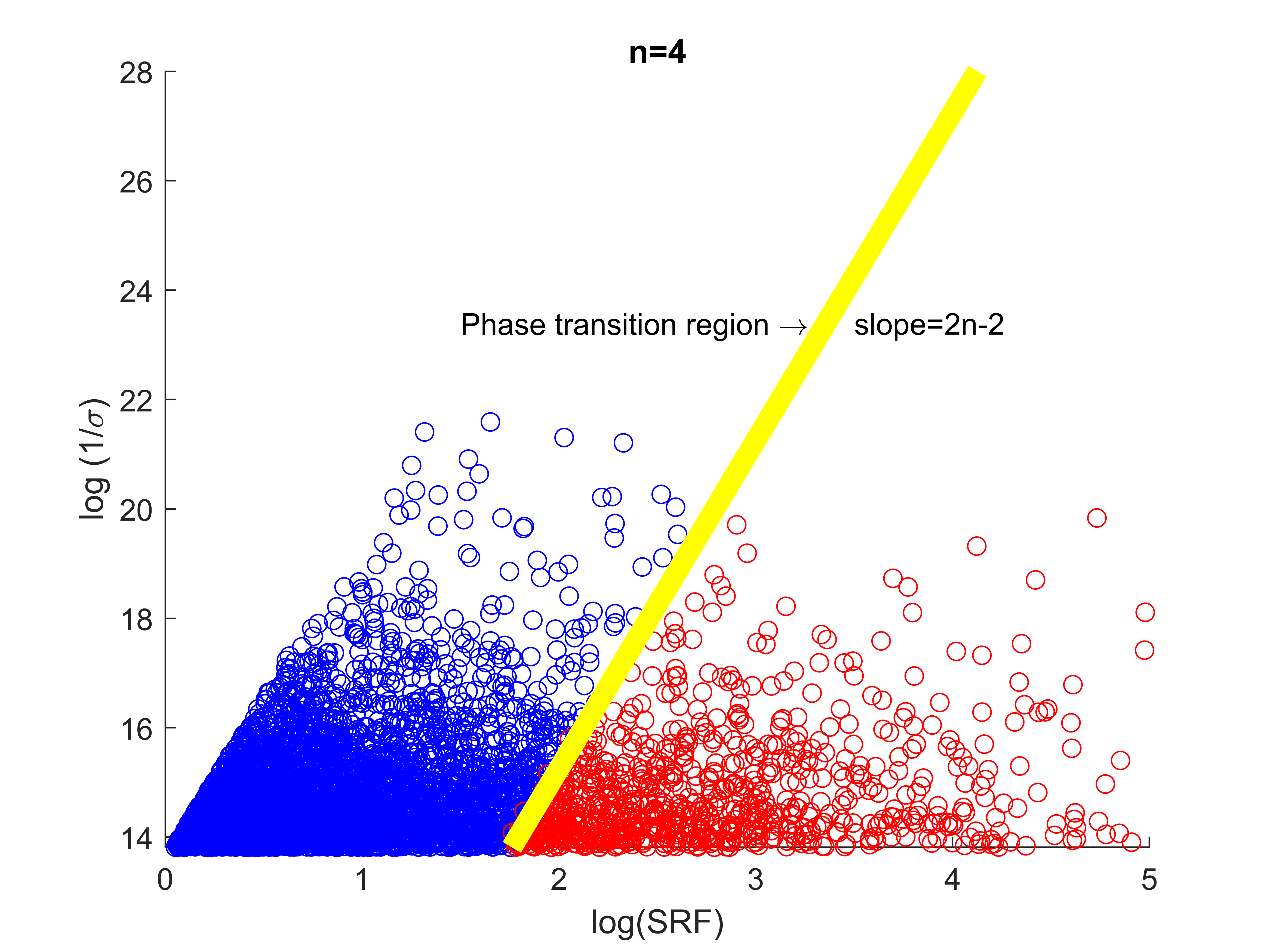}
	\caption{phase transition region}
\end{subfigure}
	\caption{Plots of the successful and the unsuccessful number detection by \textbf{Algorithm 2} depending on the relation between $\log(SRF)$ and $\log(\frac{1}{\sigma})$. (a) illustrates that two  spectra can be exactly detected if $\log(\frac{1}{\sigma})$ is above a line of slope $2$ in the parameter space. Conversely, for the same case, (b) shows that the number detection fails if $\log(\frac{1}{\sigma})$ falls below another line of slope $2$. (f) highlights the phase transition region (the yellow region) which is bounded by the black slashes in (a) and (b). (d),(e) and (f) illustrate parallel results for four spectra.}
	\label{fig:numberphasetransition}
\end{figure}

	We fix $\Omega=1$ and consider $n$ line spectra equally spaced in $\big[-\frac{(n-1)\pi}{2}, \frac{(n-1)\pi}{2} \big]$ by $d_{\min}$ with amplitudes $a_j$, and the noise level is $\sigma$. We perform 5000 random experiments (the randomness is in the choice of $(d_{\min},\sigma, y_j, a_j)$) to detect the number based on \textbf{Algorithm 2}. Figure \ref{fig:numberphasetransition} shows the results for $n=2,4$ respectively. In each case, two lines of slope $2n-2$ strictly separate the blue points (successful recoveries) and red points (unsuccessful recoveries) and in-between is the phase transition region.

For the support recovery problem, as is shown in Section \ref{section:Mainresults}, the lower and upper bounds for the computational resolution limit $\mathcal{D}_{supp}$ imply a phase transition phenomenon. Similar to the number detection, we see that in the parameter space of $\log SNR-\log SRF$, there exist two lines both with slope $2n-1$ such that the support recovery is successful for cases above the first line and unsuccessful for cases below the second. This phase transition phenomenon has been demonstrated numerically using the Matrix Pencil method in \cite{batenkov2019super}. Especially, in Section 3 of \cite{batenkov2019super}, the authors performed 15000 random experiments and showed that a line of slope $2n-1$ in the parameter space of $\log(\frac{1}{\sigma})-\log SRF$ separates the successful recoveries and the failed recoveries. Such phase transition phenomenon was also demonstrated numerically in the multi-clumps setting using the MUSIC and ESPRIT algorithms \cite{liao2016music, li2018stable, li2019super}. 
\section{Conclusion}
In this paper, we introduced two resolution limits for the number detection  and the support recovery respectively in the LSE problem. We quantitatively characterized the two limits by establishing their sharp upper and lower bounds for a cluster of spectra which are bounded by multiple of Rayleigh limit. We developed sweeping singular-value-thresholding algorithm for the number detection problem and provided a theoretical analysis. We further applied the algorithm to demonstrate the phase transition phenomenon in the number detection problem. 
The results offer a starting point for several interesting topics in the future research. First, one may extend the study to multiple clusters and design a fast and efficient algorithm to take advantage of the separation of clusters. Second, one may extend the study to higher dimensions. And finally one may extend the study to multiple measurements to gain better resolution limits. 
%And finally, a comprehensive comparison, both theoretically and numerically, of various source or spectral resolving algorithms, showing their advantages and disadvantages. 

	\section{Appendix}   
	In this Appendix, we present some inequalities that are used in this paper. We first recall the following Stirling approximation of factorial
	\begin{equation}\label{stirlingformula}
	\sqrt{2\pi} n^{n+\frac{1}{2}}e^{-n}\leq n! \leq e n^{n+\frac{1}{2}}e^{-n},
	\end{equation}
	which will be used frequently in subsequent derivation.
	
	\begin{lem}\label{numberlowerboundcalculate1}
		For $n\geq 2$, we have
		\[\frac{(n-1)^{2n-1}}{(2n-2)!}\leq \frac{\sqrt{n-1}}{2\sqrt{\pi}}(\frac{e}{2})^{2n-2}, \quad \frac{n^{2n+1}}{(2n-1)!}\leq \frac{en^2}{2\sqrt{\pi(n-\frac{1}{2})}}(\frac{e}{2})^{2n-1}.\]
	\end{lem}
	Proof: By (\ref{stirlingformula}),
	\begin{align*}
	&\frac{(n-1)^{2n-1}}{(2n-2)!}\leq \frac{(n-1)^{2n-1}}{\sqrt{2\pi}(2n-2)^{2n-2+\frac{1}{2}}e^{-(2n-2)}}=\frac{\sqrt{n-1}}{2\sqrt{\pi}}(\frac{e}{2})^{2n-2},\\
	&\frac{n^{2n+1}}{(2n-1)!}\leq \frac{n^{2n+1}}{\sqrt{2\pi}(2n-1)^{2n-1+\frac{1}{2}}e^{-(2n-1)}}\leq \frac{n^2}{2\sqrt{\pi(n-\frac{1}{2})}}(\frac{e}{2})^{2n-1}\frac{n^{2n-1}}{(n-\frac{1}{2})^{2n-1}}
	\leq \frac{en^2}{2\sqrt{\pi(n-\frac{1}{2})}}(\frac{e}{2})^{2n-1}.
	\end{align*}
	
	\medskip
	\begin{lem}\label{upperboundnumbercalculate1}
Let $\zeta(n)$ and $\xi(n-1)$ be defined as in (\ref{zetaxiformula1}).		For $n\geq 2$, we have 
		\[\Big(\frac{2\sqrt{2n-1}}{\zeta(n)\xi(n-1)}\Big)^{\frac{1}{2n-2}}\leq \frac{4.4 e}{2n-1}. \]
		
	\end{lem}
	Proof: For $n=2,3,4$, it is easy to check that the above inequality holds. Using (\ref{stirlingformula}), we have for odd $n\geq 5$,
	\begin{align*}
	&\zeta(n)\xi(n-1)= (\frac{n-1}{2}!)^2\frac{(\frac{n-3}{2}!)^2}{4}\geq \pi^2(\frac{n-1}{2})^n(\frac{n-3}{2})^{n-2}e^{-(2n-4)}\\
	= & (n-\frac{1}{2})^{2n-2}\frac{\pi^2(\frac{n-1}{2})^n(\frac{n-3}{2})^{n-2}e^{-(2n-4)}}{(n-\frac{1}{2})^{2n-2}}= \pi^2e^2(\frac{n-\frac{1}{2}}{2e})^{2n-2}\frac{(n-1)^n(n-3)^{n-2}}{(n-\frac{1}{2})^{2n-2}}\\
	=& \pi^2e^2(\frac{n-\frac{1}{2}}{2e})^{2n-2}(\frac{n-1}{n-\frac{1}{2}})^{n}(\frac{n-3}{n-\frac{1}{2}})^{n-2}\geq 0.048(e\pi)^2 (\frac{n-\frac{1}{2}}{2e})^{2n-2}, \quad \Big(\text{since $n\geq 5$}\Big)
	\end{align*}
	%Thus by $n\geq 5$,
	%\[\Big(\frac{\sqrt{2n-1}}{\zeta(n)\xi(n-1)}\Big)^{\frac{1}{2n-2}}\leq \frac{2e}{n} \Big(\frac{\sqrt{2n-1}e^2}{\pi^2}\Big)^{\frac{1}{2n-2}}\leq \frac{2.22 e}{n}. \]
	and for even $n\geq 6$,
	\begin{align*}
	&\zeta(n)\xi(n-1)= (\frac{n}{2})!(\frac{n-2}{2})!\frac{(\frac{n-2}{2})!(\frac{n-4}{2})!}{4}\geq \pi^2(\frac{n}{2})^{\frac{n+1}{2}}(\frac{n-2}{2})^{n-1}(\frac{n-4}{2})^{\frac{n-3}{2}}e^{-(2n-4)}\\
	= & (n-\frac{1}{2})^{2n-2}\frac{\pi^2(\frac{n}{2})^{\frac{n+1}{2}}(\frac{n-2}{2})^{n-1}(\frac{n-4}{2})^{\frac{n-3}{2}}e^{-(2n-4)}}{(n-\frac{1}{2})^{2n-2}}=\pi^2 e^2 (\frac{n}{2e})^{2n-2}\frac{n^{\frac{n+1}{2}}(n-2)^{n-1}(n-4)^{\frac{n-3}{2}}}{(n-\frac{1}{2})^{2n-2}}\\
	>&0.048(e\pi)^2 (\frac{n-\frac{1}{2}}{2e})^{2n-2}.   
	\end{align*}
	Therefore, for all $n\geq 5$, 
	\[\Big(\frac{2\sqrt{2n-1}}{\zeta(n)\xi(n-1)}\Big)^{\frac{1}{2n-2}}\leq \frac{2e}{n-\frac{1}{2}} \Big(\frac{2\sqrt{2n-1}}{0.048(e\pi)^2}\Big)^{\frac{1}{2n-2}}\leq \frac{2.2 e}{n-\frac{1}{2}}= \frac{4.4e}{2n-1}. \]

\begin{lem}\label{upperboundsupportcalculate1}
Let $\zeta(n)$ and $\lambda(n)$ be defined in (\ref{zetaxiformula1}) and (\ref{equ:lambda1}) respectively. For $n\geq 2$, we have 
		\[
		\Big(\frac{8\sqrt{2n}}{\zeta(n)\lambda(n)}\Big)^{\frac{1}{2n-1}}\leq \frac{5.88 e}{2n}.
		\]
\end{lem}
	Proof: For $n=2,3,4,5$, the inequality follows from direct calculation. By the Stirling approximation (\ref{stirlingformula}), we have for even $n\geq 6$,
	\begin{align*}
	&\zeta(n)\lambda(n)=\zeta(n)\xi(n-2)\leq (\frac{n}{2})!(\frac{n-2}{2})!\frac{(\frac{n-4}{2}!)^2}{4}= \pi^2(\frac{n}{2})^{\frac{n+1}{2}}(\frac{n-2}{2})^{\frac{n-1}{2}}(\frac{n-4}{2})^{n-3}e^{-(2n-5)}\\
	= & n^{2n-1}\frac{\pi^2(\frac{n}{2})^{\frac{n+1}{2}}(\frac{n-2}{2})^{\frac{n-1}{2}}(\frac{n-4}{2})^{n-3}e^{-(2n-5)}}{n^{2n-1}}=(\frac{n}{2e})^{2n-1}\frac{\pi^2e^42^2}{n^2}  \frac{(n)^{\frac{n+1}{2}}(n-2)^{\frac{n-1}{2}}(n-4)^{n-3}}{n^{2n-3}}\\
	\geq& (\frac{n}{2e})^{2n-1}\frac{4\pi^2}{en^2},
	\end{align*}
	%Thus for even $n\geq 6$,
	%\[\Big(\frac{4\sqrt{2n}}{\zeta(n)\lambda(n)}\Big)^{\frac{1}{2n-1}}\leq \frac{2e}{n+\frac{1}{2}}\Big(\frac{e^2(n+\frac{1}{2})^2\sqrt{2n}}{\pi^2}\Big)^{\frac{1}{2n-1}}\leq \frac{3.07 e}{n}.\]
	and for odd $n\geq 7$, 
	\begin{align*}
	&\zeta(n)\lambda(n)=\zeta(n)\xi(n-2)=(\frac{n-1}{2}!)^2\frac{(\frac{n-3}{2})!(\frac{n-5}{2})!}{4}\geq \pi^2(\frac{n-1}{2})^{n}(\frac{n-3}{2})^{\frac{n-2}{2}}(\frac{n-5}{2})^{\frac{n-4}{2}}e^{-(2n-6)}\\
	= & n^{2n-1}\frac{\pi^2(\frac{n-1}{2})^{n}(\frac{n-3}{2})^{\frac{n-2}{2}}(\frac{n-5}{2})^{\frac{n-4}{2}}e^{-(2n-6)}}{n^{2n-1}}= (\frac{n}{2e})^{2n-1}\frac{\pi^2e^52^2}{n^2}  \frac{(n-1)^{n}(n-3)^{\frac{n-2}{2}}(n-5)^{\frac{n-4}{2}}}{n^{2n-3}}\\
	\geq& (\frac{n}{2e})^{2n-1}\frac{4\pi^2}{n^2}. 
	\end{align*}
	Therefore, for all $n\geq 6$,
	\[
	\Big(\frac{8\sqrt{2n}}{\zeta(n)\lambda(n)}\Big)^{\frac{1}{2n-1}}\leq \frac{2e}{n}\Big(\frac{en^28\sqrt{2n}}{4\pi^2}\Big)^{\frac{1}{2n-1}}\leq \frac{2.94 e}{n}=\frac{5.88 e}{2n}.
	\]

	\begin{lem}\label{upperboundsupportcalculate2}
Let $\zeta(n)$ is defined as in (\ref{zetaxiformula1}).	For $n\geq 2$, we have 
		\[\frac{(2n)^{2n-\frac{3}{2}}}{\zeta(n)(n-2)!}\leq 2^{3n-3}e^{2n}\pi^{-\frac{3}{2}}. \]
\end{lem}
	Proof: By the Stirling approximation formula (\ref{stirlingformula}), when $n$ is odd and $n\geq 3$, we have 
	\begin{align*}
	&\frac{(2n)^{2n-\frac{3}{2}}}{\zeta(n)(n-2)!}=\frac{(2n)^{2n-\frac{3}{2}}}{(\frac{n-1}{2}!)^2(n-2)!}\leq \frac{(2n)^{2n-\frac{3}{2}}}{(\sqrt{2\pi})^3(\frac{n-1}{2})^{n}(n-2)^{n-2+\frac{1}{2}}e^{-(2n-3)}}\\
	\leq & \frac{2^{3n-\frac{3}{2}}e^{2n}}{(e\sqrt{2\pi})^3}\frac{n^{n}n^{n-\frac{3}{2}}}{(n-1)^n (n-2)^{n-\frac{3}{2}}}\leq\frac{2^{3n-3}e^{2n}}{(\sqrt{\pi})^3}
	\end{align*}
	When $n$ is even and $n\geq 4$, we have 
	\begin{align*}
	&\frac{(2n)^{2n-\frac{3}{2}}}{\zeta(n)(n-2)!}=\frac{(2n)^{2n-\frac{3}{2}}}{(\frac{n}{2})!(\frac{n-2}{2})!(n-2)!}\leq \frac{(2n)^{2n-\frac{3}{2}}}{(\sqrt{2\pi})^3(\frac{n}{2})^{\frac{n+1}{2}}(\frac{n-2}{2})^{\frac{n-1}{2}}(n-2)^{n-2+\frac{1}{2}}e^{-(2n-3)}}\\
	\leq & \frac{2^{3n-\frac{3}{2}}e^{2n}}{(e\sqrt{2\pi})^3}\frac{(n)^{\frac{n+1}{2}}(n)^{\frac{n-1}{2}}(n)^{n-\frac{3}{2}}}{n^{\frac{n+1}{2}}(n-2)^{\frac{n-1}{2}} (n-2)^{n-\frac{3}{2}}}\leq \frac{2^{3n-3}e^{2n}}{(\sqrt{\pi})^3}.
	\end{align*}
For $n=2$, the inequality follows from direct calculation. 

\begin{lem}\label{musiccalculate1}
Let $\zeta(n)$ be defined as in (\ref{zetaxiformula1}).	For $n\geq 2$, we have 
	\[
	n\Big(\frac{2n(n+1)}{\zeta(n)^2}\Big)^{\frac{1}{2n-2}}<3e.
	\]
\end{lem}
Proof: By (\ref{stirlingformula}), when $n$ is odd and $n\geq 3$, we have
\begin{align*}
&n\Big(\frac{2n(n+1)}{\zeta(n)^2}\Big)^{\frac{1}{2n-2}} = \Big(\frac{n^{2n-2}2n(n+1)}{(\frac{n-1}{2}!)^4}\Big)^{\frac{1}{2n-2}}\leq \Big(\frac{n^{2n-2}2n(n+1)}{(2\pi)^2 (\frac{n-1}{2})^{2n}e^{-2(n-1)}}\Big)^{\frac{1}{2n-2}}\\
=& 2e \Big(\frac{8n(n+1)}{(2\pi)^2 (n-1)^{2}}\frac{n^{2n-2}}{(n-1)^{2n-2}}\Big)^{\frac{1}{2n-2}}=2e(1+\frac{1}{n-1}) \Big(\frac{8n(n+1)}{(2\pi)^2 (n-1)^{2}}\Big)^{\frac{1}{2n-2}} \\
< &3e \quad \Big(\text{since $n\geq 3$ and $\frac{8n(n+1)}{(2\pi)^2 (n-1)^{2}}<1$}\Big). 
\end{align*} 
On the other hand, when $n$ is odd and $n\geq 4$, 
\begin{align*}
&n\Big(\frac{2n(n+1)}{\zeta(n)^2}\Big)^{\frac{1}{2n-2}} = \Big(\frac{n^{2n-2}2n(n+1)}{((\frac{n}{2})!(\frac{n-2}{2})!)^2}\Big)^{\frac{1}{2n-2}}\leq \Big(\frac{n^{2n-2}2n(n+1)}{(2\pi)^2 (\frac{n}{2})^{n+1} (\frac{n-2}{2})^{n-1}e^{-2(n-1)}}\Big)^{\frac{1}{2n-2}}\\
=& 2e \Big(\frac{8(n+1)}{(2\pi)^2n}\frac{n^{n-1}}{(n-2)^{n-1}}\Big)^{\frac{1}{2n-2}}=2e\Big(1+\frac{2}{n-2}\Big)^{\frac{1}{2}} \Big(\frac{8n}{(2\pi)^2n}\Big)^{\frac{1}{2n-2}} \\
< &3e \quad \Big(\text{since $n\geq 3$ and $\frac{8(n+1)}{(2\pi)^2n}<1$}\Big). 
\end{align*} 
For $n=2$, the inequality follows from direct calculation.

	\bibliographystyle{plain} 
	\bibliography{references} 
	
\end{document}